\documentclass{article}
\usepackage[utf8]{inputenc}
\usepackage{graphicx} 
\usepackage[numbers]{natbib}
\usepackage[T1]{fontenc}    
\usepackage{hyperref}       
\usepackage{amsmath,amsfonts,amssymb}
\usepackage{comment}
\usepackage[section]{placeins} 
\usepackage{multirow}
\usepackage{tabularx}
\usepackage{array}
\usepackage{pdfpages}
\usepackage{booktabs}
\usepackage{longtable}
\usepackage{pdflscape}
\usepackage{authblk}
\usepackage{siunitx}

\title{Reinforcement Learning Guides Generative Protein Language Models}
\date{}
\author[1,2]{Filippo Stocco}
\author[2,1]{Maria Artigues-Lleixà}
\author[1,3]{Andrea Hunklinger}
\author[1,2]{Michele Garibbo}
\author[4,5]{Talal Widatalla}
\author[2,6]{Marc Güell}
\author[1,2]{Noelia Ferruz\thanks{Corresponding author: \href{mailto:noelia.ferruz@crg.eu}{noelia.ferruz@crg.eu}}}

\affil[1]{Centre for Genomic Regulation, the Barcelona Institute of Science and Technology, Dr.\ Aiguader 88, Barcelona 08003, Spain}
\affil[2]{Universitat Pompeu Fabra, Barcelona, Spain}
\affil[3]{Universitat de Barcelona, Facultat de Farmàcia i Ciències de l’Alimentació, Avda.\ Diagonal 643, Barcelona 08028, Spain}
\affil[4]{Stanford University}
\affil[5]{Arc Institute}
\affil[6]{Institució Catalana de Recerca i Estudis Avançats (ICREA)}

\begin{document}
\maketitle
\begin{abstract}
Protein engineering can optimize molecules for biotechnology and therapeutics, but navigating the high-dimensional sequence landscape remains challenging. Protein language models (pLMs) have shown to to generate functional proteins far from natural sequences, yet their outputs tend to reflect prevalent properties in training data, limiting discovery of rare properties such as high catalytic activity or thermostability. Here, we introduce ProtRL, a reinforcement learning framework for pLMs that iteratively updates model parameters to maximize externally defined reward functions. Across diverse design tasks, ProtRL shifts generation toward specified objectives while maintaining sequence diversity. We demonstrate the optimization of target folds, bounded and continuous fitness predictors, and multi-objective optimization in binder design. As a proof of concept, we applied ProtRL to experimental feedback for the engineering of epidermal growth factor receptor binders. Testing fewer than 100 designed variants across the experimental campaign, ProtRL-guided optimization provided a final round in which 16 of 22 variants bound EGFR. The best variant showed a dissociation constant of 5.5 nM, representing a nine-fold improvement over wild-type EGF and higher affinity than previously reported EGF variants identified through substantially larger screening campaigns. Our code and models are publicly available at \texttt{github.com/AI4PDLab/ProtRL}.
\end{abstract}

\section*{Introduction}
Protein engineering seeks to optimize proteins for defined functional and biophysical properties, including catalytic activity, stability, selectivity, and binding affinity. However, the vast number of possible sequences makes comprehensive exploration infeasible, and experimental screening can cover only a small fraction of variants. Developing engineered protein sequences for biotechnological or therapeutic applications can take more than a decade and require investments exceeding \$500 million   \cite{wouters2024quantifying, national2025aligning}. Reducing costs, accelerating discovery-to-development timelines, and improving control over the design process are therefore central goals in modern protein engineering.

Directed evolution remains the gold standard for protein optimization. By iteratively introducing genetic diversity and selecting improved variants, it effectively performs a local, hill-climbing search through sequence space. Its impact on protein engineering has been profound \cite{arnold2017directed, mccafferty1990phage} and was recognized by the 2018 Nobel Prize in Chemistry \citep[Dr. F. Arnold,][]{nobel_chemistry_2018_pressrelease}. However, practical constraints limit access to vast regions of sequence space. For even modest proteins, the number of possible variants at Hamming distance $d=3$--$4$ from a parent sequence can already reach millions to billions, while still covering only a tiny local neighborhood. Moreover, because most random mutations are deleterious and epistasis can render high-performing combinations inaccessible, campaigns often require extensive screening with multiple rounds.

Over the past five years, protein language models (pLMs) have emerged as versatile tools enabling advances from structure prediction \cite{lin2022language} to variant fitness prediction and, particularly for autoregressive pLMs, generation of functional sequences across large regions of sequence space \cite{ruffolo2024designing, yang2025dayhoff}. These models have produced functional lysozymes, triosephosphate isomerases, and superoxide dismutases \cite{madani2023large, johnson2025computational, munsamy2024conditional}, in some cases achieving activities comparable to natural counterparts while sharing as little as 30\% sequence identity with known natural proteins \cite{romero2024novo}. These results show that pLMs can generate diverse, high-quality candidates at scale, expanding the repertoire of accessible protein families.

Despite these advances, unconstrained pLM sampling offers limited control over which functional or biophysical properties are enriched among generated sequences, as models naturally sample from their underlying training distributions and reproduce their most salient properties. This limits the ability to steer generation toward task-specific objectives, as a result, most prior work relies on extensive downstream filtering \cite{madani2023large, romero2024novo, nguyen2024sequence}. Achieving property-level control instead requires shifting the generation distribution itself toward customizable regions of sequence space, while balancing exploration and exploitation across competing objectives.

In natural language processing (NLP), an analogous challenge (steering large language models (LLMs) toward user intent or preferences rather than the most frequent patterns in their training data) is commonly addressed with reinforcement learning (RL) and preference optimization. These methods update a base model using feedback signals that score outputs as better/worse or good/bad, thereby aligning generation with an external objective. Because autoregressive pLMs share architectures and training objectives with LLMs, they are natural candidates to benefit from the same alignment toolbox. Within the rapidly advancing landscape of methods for steering pLMs \cite{stocco2025guiding}, RL-style methods are particularly appealing: they benefit from the same features that have contributed to LLM alignment without relying on human annotations.

Several methods have adopted RL for protein sequences, including DNA-binder design \citep[e.g.][]{angermueller2019model}, thermostability prediction \citep[e.g.][]{widatalla2024aligning}, binder modeling \citep[e.g.][]{wang2023self}, and alignment of large pre-trained models \citep[e.g.][]{mistani2024preference,dharuman2024mprot, blalock2025functional}. However, these efforts are often task-specific or method-specific, and the field still lacks a general, modular framework for aligning decoder-only pLMs with heterogeneous computational or experimental rewards. Here we introduce ProtRL, a modular RL alignment layer for decoder-only pLMs that supports design--score--update loops with multiple oracles and scalable training. ProtRL is directly compatible with any decoder-only pLM available through HuggingFace (\url{https://github.com/AI4PDLab/ProtRL}). 

We evaluate ProtRL across diverse classes of objectives, both \textit{in silico} and experimentally. Our results show that ProtRL enables controllable enrichment of specific folds and enzymatic functions, optimization of continuous metrics such as ESM-1v \cite{meier2021language} and ProteinMPNN \cite{dauparas2022robust} log-likelihoods, and integration of learned reward models trained on activity data. We further demonstrate its applicability by designing epidermal growth factor (EGF) variants and evaluating them by bio-layer interferometry (BLI), achieving a nine-fold improvement over wild-type EGF within fewer than 100 experimentally tested designs and surpassing previously reported engineered variants obtained through substantially larger screening efforts \cite{lahti2011engineered,cochran2014mutant}. In addition to the framework, we provide practical guidance on reward construction, oracle selection, and diagnosis of common failure modes. Overall, ProtRL provides a modular route for optimizing user-defined protein properties using computational or experimental feedback while preserving the generative diversity of pre-trained pLMs.

\begin{figure}
    \centering
        \includegraphics[ page=1, width=\linewidth, trim=2cm 35cm 5cm 15cm, clip]{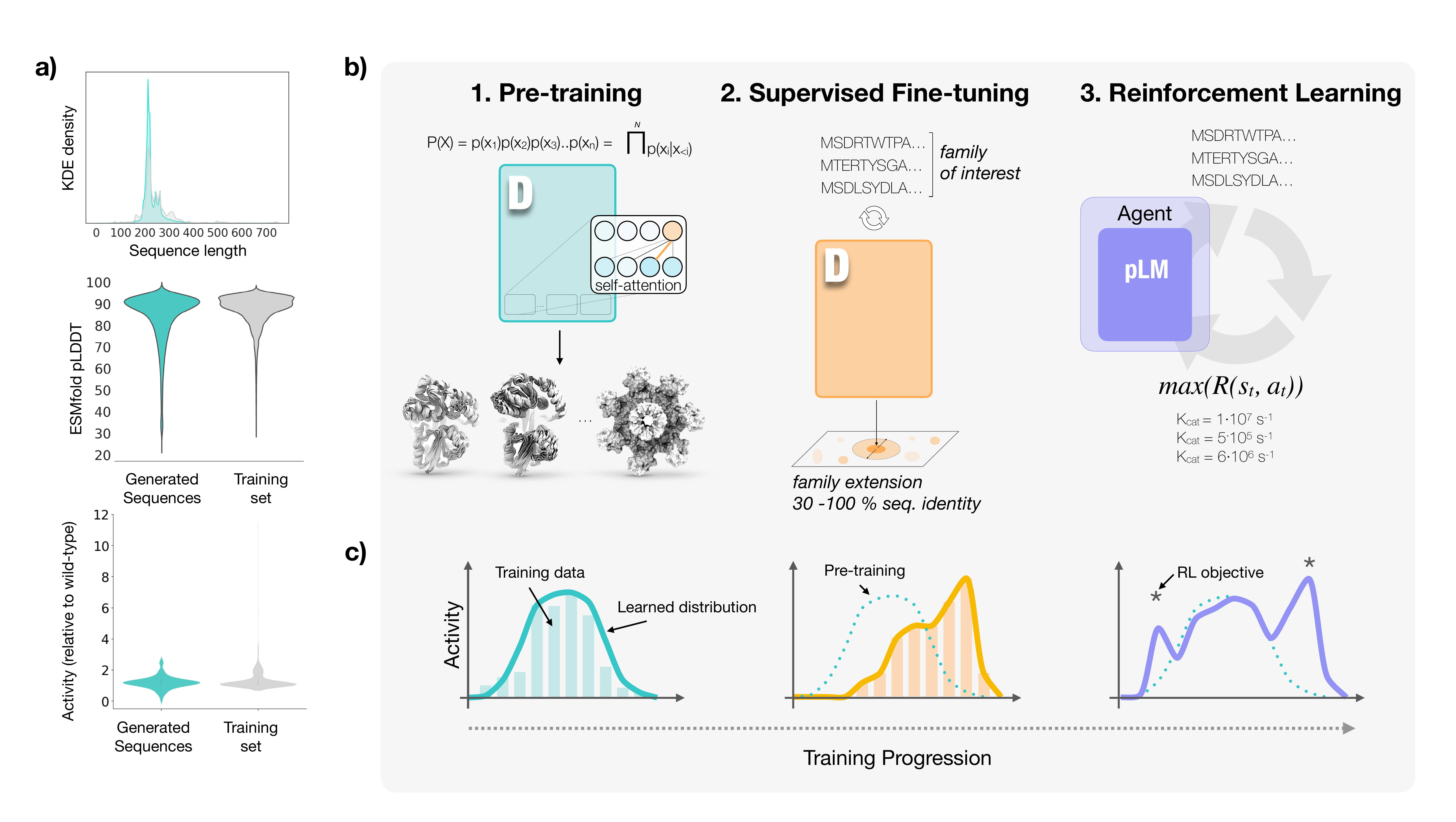}
    \caption{\textbf{Protein language models reproduce dominant properties of their training distributions and can be redirected by reinforcement learning.} 
    \textbf{a,} Unconstrained generation from protein language models tends to recapitulate properties of the training data, including sequence length, predicted structural confidence (pLDDT), and experimentally measured activity relative to the wild type. This limits access to rare but desirable regions of sequence space, such as variants with unusually high activity, stability, or other engineering-relevant properties. 
    \textbf{b,} In ProtRL, we adapt pLMs through successive stages of pre-training (PT), supervised fine-tuning (SFT), and reinforcement learning (RL) or preference optimization. During pre-training, the model learns a broad, high-dimensional probability distribution over amino-acid sequences by autoregressively modeling large, unlabeled sequence datasets. SFT uses additional labeled data to bias generation toward a specific region of sequence space, such as a target protein family, but remains agnostic to functional or biophysical properties not explicitly encoded in the fine-tuning data. In contrast, RL treats the model as a policy and updates it to maximize an explicit reward function, such as enzymatic activity, thermostability, expression level, or binding affinity. A Kullback--Leibler (KL) divergence term constrains the learned policy to remain close to the reference model, preserving knowledge acquired during pre-training while enabling exploration of higher-reward regions of sequence space.}
    \label{fig:fig1}
\end{figure}

\subsection*{pLMs do not transcend the limits of their underlying training distributions}

Autoregressive pLMs define an explicit probability distribution over protein sequences,
\begin{equation}
p_{\theta}(x_{1:T})=\prod_{t=1}^{T} p_{\theta}(x_t \mid x_{<t}),
\end{equation}
where $x_t$ denotes the amino acid at position $t$. As $p_{\theta}$ is typically fit by maximum likelihood on a finite training corpus, unconstrained generation tends to reproduce sequence-level and coarse structural properties represented in the training dataset, such as length, amino-acid composition, and predicted fold-related features, rather than systematically enriching rare, high-fitness regions of sequence space.

To quantify this effect in an enzyme-family setting, we generated 40{,}000 sequences with ZymCTRL, an autoregressive pLM conditioned on EC labels \cite{munsamy2022zymctrl}. The generated sequences recapitulated properties of the corresponding training subset, including the sequence-length distribution and predicted structural confidence (pLDDT; Fig.~\ref{fig:fig1}a), as well as additional structural features (Fig. \ref{fig:SI_alphaamylase}). This illustrates that pLMs can faithfully sample prominent modes of the training distribution for properties that are strongly encoded in sequence statistics and structural priors.

Functional phenotypes, however, are higher-order properties that depend on expression, folding, stability, dynamics, and precise active-site or interface geometry. Consequently, matching the training distribution in sequence length or predicted structural confidence does not imply matching the training distribution in catalytic efficiency, binding affinity, or expression. Indeed, previous experimental studies have shown that generative pLMs can produce functional proteins, including sequences highly divergent from known natural proteins \cite{madani2023large, johnson2025computational, munsamy2024conditional, romero2024novo}. However, especially as generated sequences move farther from natural homologs, functional outcomes often shift below the range observed from natural training sequences (positive controls), with reduced expression, lower activity, or lower hit rates. Thus, pLMs provide powerful generative priors for sampling diverse and often functional proteins, but unconstrained sampling alone is not an optimization strategy for enriching rare, high-fitness variants. This motivates approaches that explicitly steer generation toward desired properties.

\subsection*{Reinforcement learning guides pLMs towards specific properties}
RL provides a principled framework for updating pre-trained generative models using scalar rewards or preference signals while regularizing against excessive deviation from the original distribution. In practice, alignment is commonly implemented using policy-gradient methods such as Proximal Policy Optimization (PPO), its group-relative variant (GRPO), or direct preference learning approaches such as Direct Preference Optimization (DPO). 

Here, we cast protein sequence design as an RL problem in which the protein language model defines a policy $p_\theta(y|x)$ that generates sequences \textit{y} conditioned on an input context $x$. The goal is to update the parameters $\theta$ to maximize expected reward:

\begin{equation}
\theta^\star \;=\; \arg\max_{\theta}\; \mathbb{E}_{x \sim D}\Big[\, \mathbb{E}_{y \sim p_\theta(\cdot \mid x)}\big[ R(x,y) \big] \Big]
    \label{reinforcement learning}
\end{equation}

At each iteration, the pLM generates a batch of diverse sequences. An oracle (predictor, physics-based metric, or experiment) assigns each sequence a reward $(R(x,y))$. These sequence--reward pairs are then used to update the pLM, biasing subsequent generations toward higher-value regions of sequence space (Fig. \ref{fig:fig1}b). We refer to \cite{stocco2025guiding}  and Methods \ref{RL_expl} for in-depth technical details.

Here, we instantiate and benchmark this framework using ZymCTRL, an enzyme language model; however, ProtRL is applicable to any autoregressive pLM, integrating seamlessly with standard HuggingFace \cite{wolf2019huggingface} generation and training pipelines, including multi-node and multi-GPU configurations.
We evaluate ProtRL across diverse objective classes: (a) steering toward a desired fold, (b) optimizing bounded objectives such as label fidelity, (c) optimizing unbounded continuous scores and experimental proxies, (d) multi-objective design of real-world binders, and (e) alignment using experimental feedback for the optimization of binding affinity. Throughout, we emphasize practical considerations to facilitate reproducible integration into protein-engineering workflows.

\subsection*{ProtRL steers generation toward a desired fold}
Carbonic anhydrases (CAs; EC~4.2.1.1) comprise multiple phylogenetically unrelated structural classes ($\alpha, \beta, \gamma, \delta, \zeta, \eta, \theta, \iota$) that catalyze the same reaction \cite{banerjee2016origin}. When conditioned on E.C.~4.2.1.1, the pre-trained base ZymCTRL model primarily generates $\alpha$- and $\beta$-class CAs, with a bias toward $\beta$, reflecting the training set composition (Fig. \ref{fig:fig2}). To actively steer ZymCTRL towards generating the target fold, we used structural similarity to a reference $\alpha$-carbonic anhydrase ($\alpha$-CA; PDB 2VVB) as the oracle signal. Structures of generated sequences were predicted with ESMFold \cite{lin2022language}, and each prediction was assigned to a structural family by superimposing the two proteins structures and computing its TM-score against the reference (Methods). TM-scores range from 0 to 1, with values above 0.5 generally indicating a shared fold topology \cite{zhang2004scoring,xu2010significant}. At each iteration, we generated 200 sequences and assigned higher rewards to sequences with higher TM-scores relative to the $\alpha$-CA reference using GRPO.

Interestingly, when optimizing TM-score alone, the model progressively increased sequence length (see Fig. \ref{fig:SI_tmscoreGRPO}), a clear example of reward hacking that has been widely observed in other optimization settings \cite{krakovna2020specification}. In this case, the model exploits a "shortcut" that inflates the reward without aligning with the underlying design objective, producing sequences that are unlikely to be biologically meaningful. Incorporating an explicit alignment-length term into the reward function stabilized optimization (Methods, Table \ref{Table:SI_loss_func}) and enabled rapid convergence toward the target fold (Fig. \ref{fig:fig2}a). Across 20 iterations, sequence diversity and quality were maintained. MMseqs \cite{steinegger2017mmseqs2} clustering at 80\% sequence identity revealed no collapse in diversity (Fig. \ref{fig:fig2}a and Fig. \ref{fig:SI_tmscoreGRPO}). Together, these results demonstrate that ProtRL can strongly bias fold outcomes while preserving sequence diversity, but also highlight the importance of carefully designing reward functions.

\subsection*{ProtRL optimizes categorical objectives}
Public enzyme databases are highly imbalanced: some EC labels have very large sequence counts, whereas many have only one or a few representatives \cite{munsamy2024conditional}. This imbalance is reflected in limited conditional generation for underrepresented enzymatic families. For example, when prompted with the pancreatic ribonuclease EC label (EC~4.6.1.18), only $\sim10\%$ of ZymCTRL generations are assigned the correct label by CLEAN \cite{yu2023enzyme} (Fig. \ref{fig:SI_tmscoreGRPO}).
To increase label fidelity, we used CLEAN as an oracle, optimizing the fraction of sequences classified as EC~4.6.1.18 and proximity in CLEAN embedding space to the target cluster center (Methods). To prevent drifts in sequence length, we applied a length-based discount (Methods, Table \ref{Table:SI_loss_func}). Within two iterations, $\sim$60\% of sequences were labeled as EC~4.6.1.18 (Fig. \ref{fig:SI_tmscoreGRPO}). However, this improvement initially reduced structural plausibility, requiring us to adopt a multi-term reward objective that additionally encouraged high pLDDT values (Methods, Table \ref{Table:SI_loss_func}). Joint optimization of this multi-term reward objective increased average pLDDT from $\sim$40 to $\sim$80 and, indirectly, improved TM-score to a reference ribonuclease (PDB 11BA) (Fig. \ref{fig:SI_tmscoreGRPO}). 

\subsection*{ProtRL optimizes unbounded, continuous objectives and experimental proxies}
\begin{figure}
    \centering
    \includegraphics[ page=2, width=\linewidth, trim=2cm 32cm 20cm 32cm, clip]{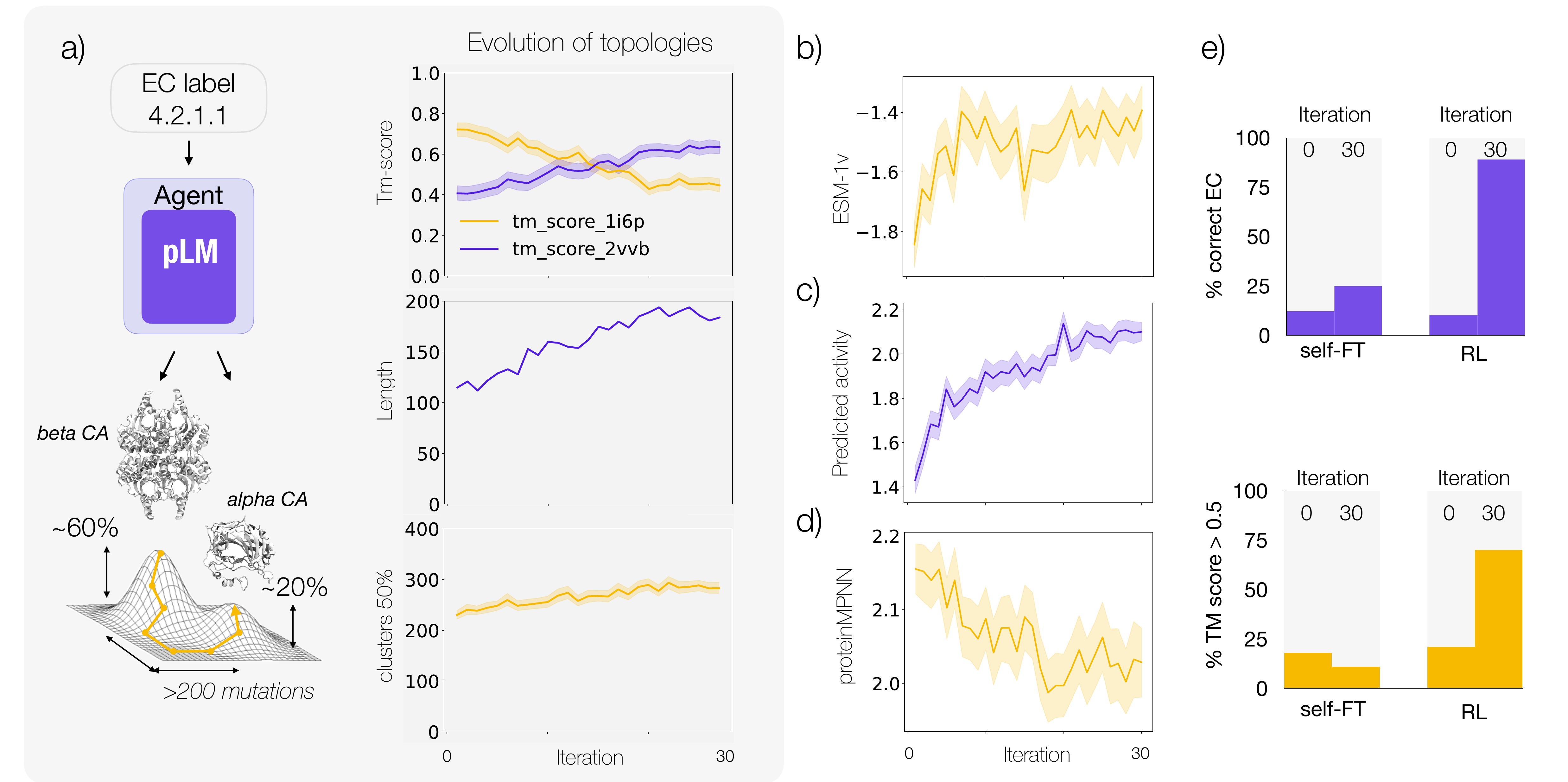}
    \caption{\textbf{ProtRL enables property-directed generation across diverse objective classes.} (a) Fold/topology steering. ProtRL is applied to an EC-conditioned pLM (example: EC 4.2.1.1) and rewarded for adopting a target topology while avoiding an off-target fold. The right panels report the evolution across iterations of (top) TM-scores to the target (purple) and off-target (orange) reference structures, together with (middle) sequence length and (bottom) functional/label-consistency score used as an auxiliary constraint. (b) Continuous oracle optimization. Iterative ProtRL improves the ESM-1v log-likelihood of generated sequences over iterations. (c) Learned experimental proxy. ProtRL optimizes predicted specific activity using a reward model trained on Protein Engineering Tournament measurements (SAPI predictor). (d) Structure-aware likelihood optimization. ProtRL increases ProteinMPNN log-likelihood across iterations. (e) RL vs. recursive self fine-tuning. Comparison between self-fine-tuning on self-generated data (self-FT) and ProtRL shows improved rare-label fidelity (top: \% sequences classified with the correct EC label) and improved fold enrichment (bottom: \% sequences with TM-score > 0.5) with ProtRL.}
    \label{fig:fig2}
\end{figure}

The objectives described above are bounded (e.g., a fraction cannot exceed 100\%). Thus, we next asked whether ProtRL can optimize continuous, effectively unbounded scores beyond the limits explored within the training distribution. We constructed reward functions to increase sequence/structure log-likelihoods under ESM1v \cite{meier2021language} and ProteinMPNN \cite{dauparas2022robust} (Methods), two metrics that have been reported to correlate with experimental outcomes in protein engineering \cite{johnson2025computational}. For both carbonic anhydrase (EC~4.2.1.1) and chorismate mutase (EC~5.4.99.5), ProtRL yielded sustained improvements in these likelihood-based scores (Fig. \ref{fig:fig2} b,d, and Fig. \ref{fig:SI_SAPIGRPO}).

To connect optimization more directly to experimental activity signals, we used the $\alpha$-amylase dataset from the Protein Engineering Tournament \cite{armer2025results} and trained a sequence-only predictor of the Specific Activity Performance Index (SAPI) (Methods). Specifically, we fine-tuned ESM1v \cite{hu2022lora, schmirler2024fine} on $\alpha$-amylase sequences paired with their corresponding SAPI values, and then used the resulting regressor as the oracle for ProtRL (Methods). This procedure improved the predicted quality of generated sequences, with predicted specific activity increasing steadily across iterations (Fig.~\ref{fig:fig2}c and Fig. \ref{fig:SI_SAPIGRPO}), demonstrating that ProtRL can successfully incorporate learned reward models derived from experimental measurements.

Given these results, we asked whether standard fine-tuning, the most widely used technique to condition pLMs \cite{munsamy2022zymctrl, madani2023large, ruffolo_design_2024, ivancic_discovery_2025}, could achieve similar improvements across metrics. We defined a \emph{self-fine-tuning} (self-FT) strategy in which pLMs were iteratively fine-tuned on the top-performing sequences generated in the previous batch. At each cycle, 2,000 sequences were generated and ranked by a predefined oracle, and the top 10\% were then used to fine-tune the model (Methods). This procedure was repeated for 30 iterations to enable direct comparison with RL-based alignment. When applied to the same objectives as ProtRL, self-FT showed limited steering capacity. For $\alpha$-carbonic anhydrase (CA) topology enrichment, incorporation of a length term preserved the canonical sequence length (260 aa; Table \ref{Table:SI_loss_func}), but the fraction of $\alpha$-like generations reached only $\sim$20\% after 30 iterations (Fig. \ref{fig:SI_SFT}a). In addition, the best TM-score declined over training (0.97 at iteration~0 versus 0.71 at iteration~30), and longer runs led to model collapse and redundancy (Fig. \ref{fig:SI_SFT}d), consistent with prior reports \cite{shumailov2024ai}. For EC~4.6.1.18, self-FT increased the fraction of CLEAN-assigned sequences to at most $\sim$30\% before performance began to degrade. Despite these modest objective gains, sequence diversity remained high throughout training (Fig. \ref{fig:SI_SFT}d). Given the greater computational cost of fine-tuning and its inferior performance relative to ProtRL (Fig. \ref{fig:fig2}e), we concluded that, under these conditions, it performs worse than RL, in agreement with recent findings \cite{xiong2025proteinguide}.

\subsection*{Multi-objective design of protein binders}
We next evaluated ProtRL in a practical design setting: multi-objective optimization of protein binders against three targets: epidermal growth factor receptor (EGFR), CD80 receptor, and Nipah virus' glycoprotein G. For each target, we initialized from a task-adapted model obtained by supervised fine-tuning on known binders augmented with sequence variants. Variants were generated either by introducing random mutations around a wild-type binder or by retrieving homologous sequences via sequence similarity search and fine-tuning on the resulting family (Methods). For the Nipah target, we additionally incorporated known antibody binders and variants thereoff (Methods).

\begin{figure}[ht]
    \centering
    \includegraphics[ page=3, width=\linewidth, trim=8cm 12cm 1cm 25cm, clip]{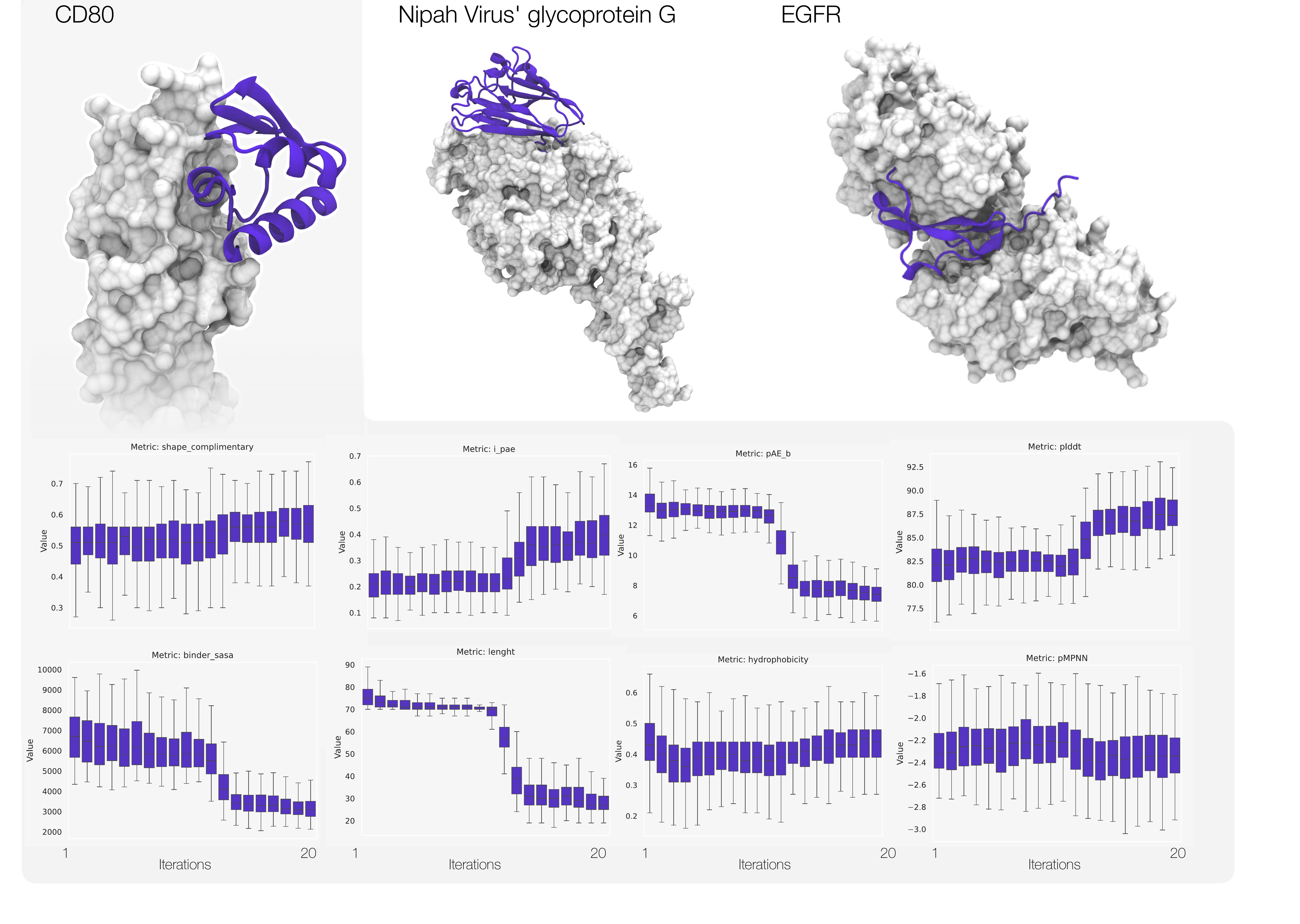}
    \caption{\textbf{Multi-objective ProtRL optimization for binder design against EGFR, CD80, and Nipah virus' glycoprotein G.} Representative AlphaFold-predicted complexes for three targets (target shown as grey surface; designed binder shown in purple) obtained after multi-objective ProtRL optimization. The lower panels track the evolution of individual reward components across iterations, illustrating consistent directional optimization under a composite objective for CD80 binders. Metrics include interface- and complex-quality terms (e.g., interface PAE, interface pTM), binder confidence/quality (e.g., binder pLDDT, pMPNN), geometric/interface complementarity terms (e.g., shape complementarity and solvent-accessible surface area (SASA)), and regularizers/penalties (e.g., length control). Together, these trends indicate effective multi-objective alignment of sequence generation with heterogeneous binder-design constraints across targets.}
    \label{fig:fig3}
\end{figure}

Starting from these target-adapted priors, we performed iterative ProtRL using a multi-term reward that captured the desired binding-related objectives (Methods). Across targets, we observed consistent directional changes of the optimized metrics over iterations, indicating effective alignment of generation with the specified multi-objective reward (Fig. \ref{fig:fig3}, Table \ref{Table:SI_loss_func}, Methods).

\subsection*{ProtRL enables the experimental design of low-nanomolar protein binders}
We evaluated ProtRL in an experimental binder-engineering campaign targeting the epidermal growth factor receptor (EGFR). As a transmembrane protein central to cell growth and a key driver in various carcinomas, EGFR is a high-priority target for protein engineering \cite{normanno2006epidermal, buch2013computational}. Traditional efforts have relied on monoclonal antibodies or variants of the endogenous ligand, epidermal growth factor (EGF). Here, we aimed to engineer novel EGF variants using ProtRL, focusing on assessing the framework's efficacy in real-world engineering workflows.

While ZymCTRL is an enzyme-specific pLM, we hypothesized that ProtRL could effectively guide its generation toward binder design by leveraging specific sequence motifs. We first performed supervised fine-tuning (SFT) on ZymCTRL using a curated dataset of 500 EGF-homologous sequences retrieved via BLAST \cite{shiryev2007improved}, prepended with the artificial EC label 1.3.3.18 (Methods, Fig. \ref{fig:fig1}). Experimental characterization was performed using an automated expression and bio-layer interferometry (BLI) pipeline, in which cell-free-expressed binders were assayed against a concentration series of EGFR and binding was monitored label-free as time-resolved optical shifts during association and dissociation to estimate $K_\mathrm{D}$, $k_{\mathrm{on}}$, and $k_{\mathrm{off}}$ (Methods). As an initial baseline, we sampled and characterized six designs from this SFT model. Three variants exhibited EGFR binding with $K_\mathrm{D}$ of 44.4 nM, 54.9 nM, and 61.0 nM, comparable to the wild-type EGF control measured at 47.4 nM under identical conditions (Fig.~\ref{fig:fig4}d, \ref{table:sequences_EGFR}).

Building on this baseline, we initiated a multi-stage alignment campaign using ProtRL with Ranked Direct Preference Optimization (rDPO). In a second round, the SFT model was aligned using a composite dataset comprising our Round 1 experimental results and EGFR data available from a previous online binder competition on EGFR \cite{cotet2025crowdsourced}. This yielded a significant improvement, with 9 of 21 tested sequences showing binding and the best affinity reaching 12.3 nM (Table \ref{table:sequences_EGFR}), corresponding to a 3.9-fold reduction in Kd relative to the wild-type.

Subsequent iterations (Rounds 3.1 and 3.2, Fig. \ref{fig:comparison_rounds}) revealed the high sensitivity of the RL process to hyperparameter selection and reference model anchoring. Specifically, an attempt to use low $\beta$ with the Round 2 model as a starting point (Round 3.1) resulted in a drastic reduction in binding hit rates (Methods). Similarly, setting the reference model to the original pre-trained pLM rather than the SFT model (Round 3.2) likely caused the generation to drift significantly from the target EGF fold (Table \ref{table:sequences_EGFR}, Figure \ref{fig:comparison_rounds}). A critical mechanistic insight gained from these iterations was the impact of data heterogeneity. We observed significant discrepancies between affinities reported in public databases and our experiments, suggesting that conflicting reward signals from different experimental environments may severely hinder the RL process. 

Consequently, we hypothesized that a successful RL-driven protein engineering campaign should follow a reward signal internally consistent and matched to the precise selection environment where the model will be deployed, and adhered to using the Round 1 model for the KL-reference model. Therefore, we performed a final alignment round (Round 3.3) using Group Relative Policy Optimization (GRPO) and relied on a reward signal derived exclusively from our internal sequences tested in the previous rounds (Methods). We identified which  \textit{in silico} metrics best discriminated binders from non-binders in the experimental data. This enabled the implementation of filters that were subsequently applied to prioritize designs for downstream testing (Figure \ref{fig:SI_filters}).

Specifically, after this alignment, we generated 10,000 sequences, of which 2,000 were selected from among the top candidates with the highest implicit reward (defined as the difference between the log likelihood of the updated model and the reference model (Equation~\ref{implicit_reward})), and the lowest perplexity, computed as the exponential of the negative average sequence log-likelihood. The top 22 sequences passing the selected structure-based filters were advanced for experimental testing. This strategy substantially improved both hit rate and affinity, yielding 16 binders, several of which reached the low-nanomolar range. Notably, two variants exhibited $K_\mathrm{D}$ values of 9.8 and 5.5 nM, both lower than m28, the strongest EGFR-binding EGF variant previously reported in the literature \cite{lahti2011engineered} and included as a control in our experimental campaign (Fig.~\ref{fig:fig4}d; Table \ref{table:sequences_EGFR}). For comparison, m28 was isolated by yeast-display directed evolution from a first-generation randomly mutagenized EGF library containing approximately $1.2 \times 10^7$ transformants, followed by multiple rounds of flow-cytometric selection for soluble EGFR binding, whereas our full campaign tested 89 variants.

A notable feature of this campaign is that ProtRL was not restricted to fixed-length optimization. Across the experimental campaign, binding variants ranged from 49 to 63 amino acids, indicating that the autoregressive model could explore insertions and deletions rather than only substitutions around a fixed scaffold. In the final round, tested designs ranged from 50 to 54 amino acids, and the 16 binders spanned the same range, whereas WT EGF and m28 are 53 amino acids. Thus, the strongest binders emerged from a search process that allowed length variation, with the two best final-round variants measuring 52 and 51 amino acids and reaching $K_\mathrm{D}$ values of 9.8 and 5.5 nM, respectively.

The final designs also did not collapse onto near-identical homologs. Instead, nearest-neighbor identity to known sequences decreased from 87.2\% in round 1 to 77.9\% in the final round (Fig.~\ref{fig:fig4}e), indicating that affinity improvements were accompanied by exploration of more diverse EGF-like sequences. Together, these results show that ProtRL can use sparse experimental feedback to enrich high-affinity binders while preserving both sequence diversity and length flexibility.

\subsection*{Top binder reveals mutations at established EGF–EGFR recognition hotspots}
\begin{figure}
    \centering
    \includegraphics[ page=4, width=\linewidth, trim=14cm 3cm 16cm 39cm, clip]{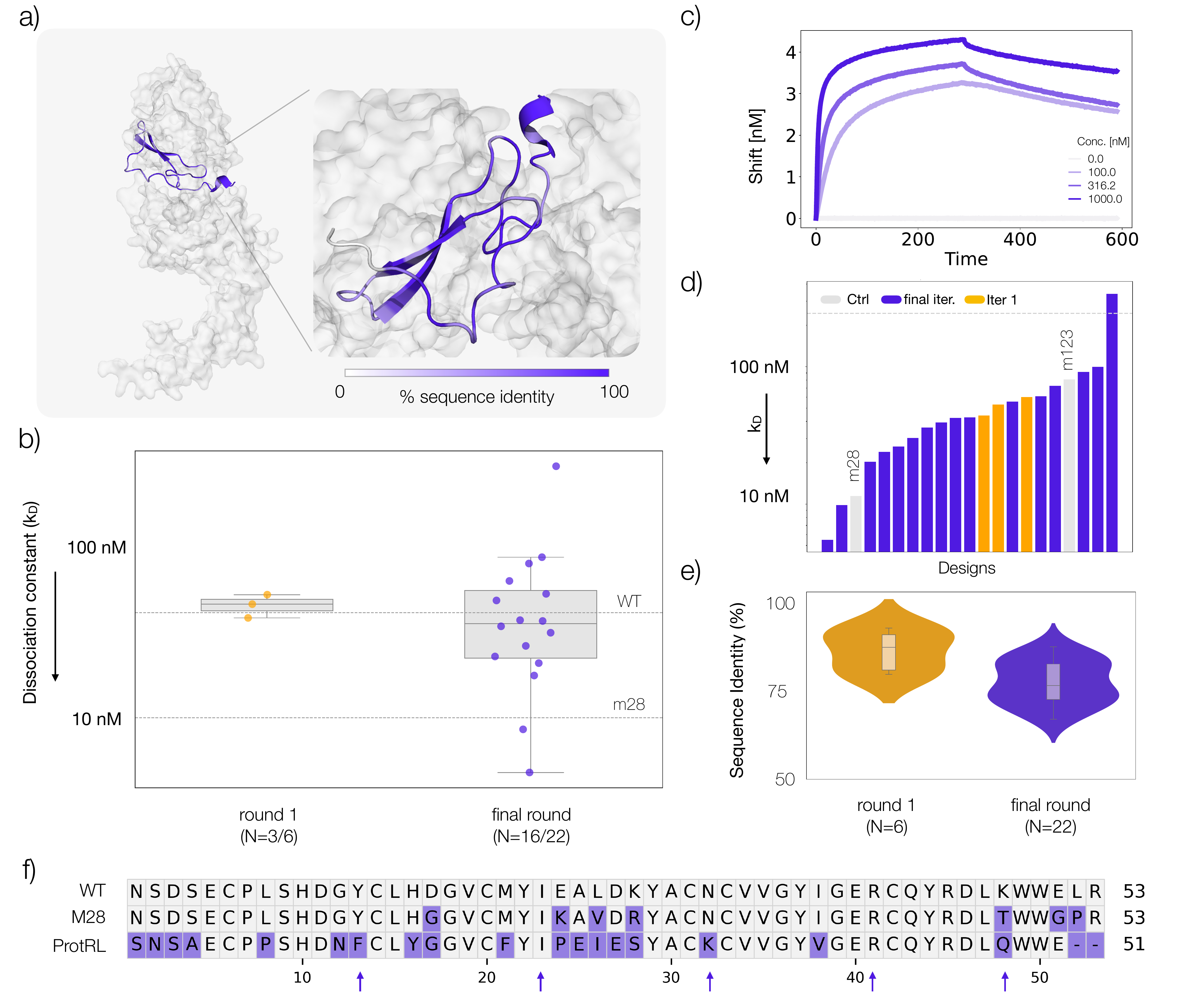}
    \caption{\textbf{Experimental EGFR binder optimization with ProtRL.}\\
(a) AlphaFold-predicted complex of the top engineered binder bound to EGFR (P00533, grey surface). The engineered binder is shown in purple, with residues colored by percent sequence identity to the strongest literature-reported EGFR binder (color bar, 0--100\%), highlighting regions of divergence while preserving the overall binding mode. (b) Distribution of measured dissociation constants ($K_\mathrm{D}$) for binders identified in the initial SFT baseline round and in the final ProtRL round. Dashed reference lines indicate the wild-type EGF and m28 controls, showing that final-round ProtRL designs are enriched for tighter binders. (c) Representative bio-layer interferometry (BLI) traces for the top engineered binder across a concentration series, showing the time-resolved signal shift used to estimate binding kinetics. (d) Measured $K_\mathrm{D}$ values for experimentally tested designs from the baseline/control set and successive ProtRL iterations. ProtRL progressively enriches for tighter binders, yielding variants that outperform the WT reference and the literature-derived m28 control under the assay conditions used here. (e) Violin plots showing the distribution of sequence identity to known homologs or nearest-neighbor sequences identified by BLAST, indicating that affinity improvements are accompanied by exploration of more diverse sequence space. (f) Sequence comparison of WT EGF, m28, and the best ProtRL-designed binder. Residues that differ from WT are highlighted in purple, and arrows indicate key binding residues reported in the literature. The ProtRL-optimized sequence is substantially more divergent than m28 and shorter than WT EGF.}
    \label{fig:fig4}
\end{figure}

The top ProtRL binder differs from both WT EGF and m28 by 18 substitutions and is two residues shorter, corresponding to 33 identical positions over the 51-residue designed binder sequence, or 64.7\% identity, and has a 76.6\% identity to any other sequence present in ClusteredNR. For comparison, m28 differs from WT EGF by seven substitutions and has the same length as WT, with an 86.8\% sequence identity (Fig. \ref{fig:fig4}f). 

Sequence alignment showed that the highest-affinity ProtRL binder retained some established features of the EGF--EGFR interface while modifying others. Residues Ile23, Arg41 and Leu47, previously implicated in receptor binding by structural and mutagenesis studies, were conserved in the optimized sequence ~\cite{ogiso2002crystal,sanders2013molecular} (Fig. \ref{fig:fig4}f). By contrast, Tyr13 was replaced by Phe, preserving an aromatic side chain while removing the hydroxyl group and Asn32 by Lys. Interestingly, the latter substitution had been linked to receptor-binding activity and predicted to contribute favorably to EGFR-ligand binding, and ProtRL selected N32K in a shorter, highly divergent binder, together with other substitutions, while retaining high affinity.

Next, we used AlphaFold ~\cite{abramson2024accurate} to predict the three-dimensional structure of the engineered binder and performed structural analysis of the predicted complex. This analysis revealed that the engineered binder forms 18 hydrogen bonds with EGFR, compared with 15 hydrogen bonds for m28, and 21 hydrophobic contacts versus 19, indicating a more extensive interaction interface (Fig.~4a). In addition, the complex displays improved structural complementarity, consistent with the observed affinity gains.

Together, these results show that ProtRL can incorporate experimental feedback to rapidly improve binding affinity while continuing to explore diverse regions of sequence space, yielding interpretable, protein-engineering-relevant mutation proposals.


\subsection*{Implicit reward as interpretability tool}
Rafailov et al. showed that DPO implicitly induces a token-level reward function even when preference signals are provided at the sequence level \cite{rafailov2024r}. This yields a principled form of credit assignment, enabling identification of tokens that drive preference. In its per-step form, the implicit reward for choosing action (token) $a$ in state $s$ is

\begin{equation}
    r(s,a) \;=\; \beta \log \pi_{\theta}(a \mid s)\;-\;\beta \log \pi_{\text{ref}}(a \mid s),
    \label{implicit_reward}
\end{equation}

where $\pi_{\theta}$ is the aligned policy and $\pi_{\text{ref}}$ is the reference policy (here, the pre-trained ZymCTRL). Practically, Eq.~\ref{implicit_reward} can be computed per token from log-likelihoods in a single forward pass, providing an amino-acid–level resolution.

\begin{figure}
    \centering
    \includegraphics[ page=5, width=\linewidth, trim=2cm 22cm 5cm 39cm, clip]{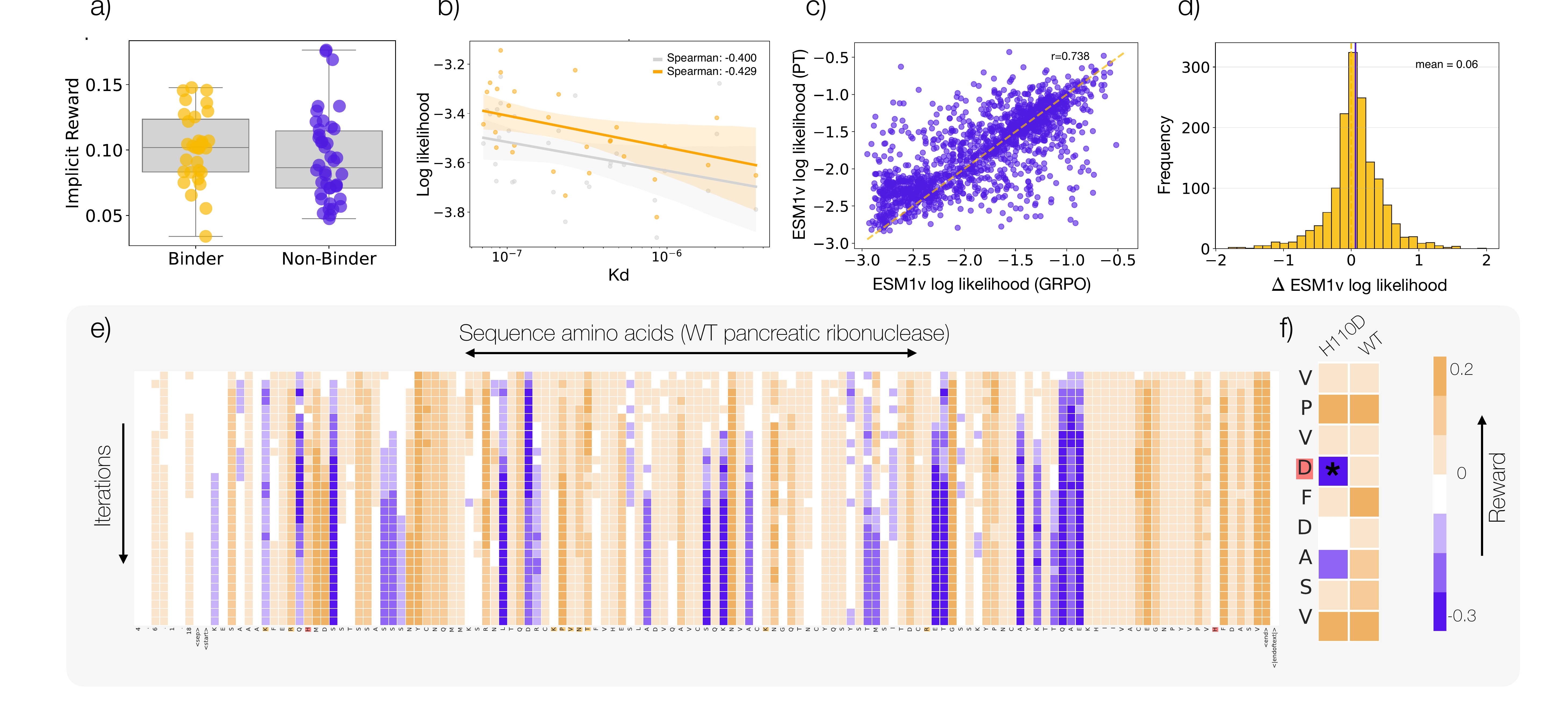}
    \caption{\textbf{Implicit token-level reward offers an interpretable signal.} (a) Relationship between experimental binding affinity ($K_\mathrm{D}$) and model log-likelihood for engineered ribonuclease variants, showing that likelihood-based signals only partially track the experimental objective (Spearman correlations reported). (b) Agreement between the implicit preference reward and an external oracle: per-sequence implicit reward correlates with ESM1v log-likelihood (Spearman $\rho$ shown), indicating that the aligned policy shifts probability mass toward sequences favored by the objective while remaining compatible with protein-likeness priors. (c) Distribution of the change in ESM1v log-likelihood induced by alignment ($\Delta$ ESM1v log-likelihood; mean indicated), summarizing the typical magnitude of the shift, highlighting that we did not observe strong evidence of substantial forgetting. (d) Token-level implicit reward $r_s$ mapped along the wild-type pancreatic ribonuclease sequence across ProtRL iterations ($\beta$ = 0.1), visualized as a heatmap that highlights positions most responsible for preference improvement. (e-f) Zoom-in on catalytic-site residues: mutating the catalytic histidine produces a strong local drop in implicit reward without broadly perturbing the surrounding neighborhood signal, illustrating residue-level attribution.}
    \label{fig:fig5}
\end{figure}

We applied this analysis to pancreatic ribonuclease (EC~4.6.1.18) under joint CLEAN and pLDDT reinforcement. Visualizing token-wise rewards as heatmaps across iterations (Fig. \ref{fig:fig5}; Fig. \ref{fig:SI_mutations_i_reward}) revealed localized reward shifts at specific positions, highlighting residues whose mutations most strongly affected the learned objective. In this sense, the implicit reward from ProtRL offers token-level attribution signals: it highlights residues that the aligned model treats as relevant to the oracle-defined constraints, generating residue-level actionable hypotheses for protein engineering.

We further analyzed the relationship between sequence log-likelihood and binding affinity ($K_\mathrm{D}$) under both the reference model (EGF fine-tuned ZymCTRL) and the aligned model. Alignment increased the correlation between model likelihood and experimental affinity: the Spearman correlation between log-likelihood and $K_\mathrm{D}$ improved from -0.400 ($p=2.59\times10^{-2}$) under the reference model to -0.429 ($p=1.61\times10^{-2}$) after alignment, while Pearson correlation showed a modest change from -0.217 ($p=2.42\times10^{-1}$) to  -0.229 ($p=2.16\times10^{-1}$). This shift suggests improved consistency between model preference and experimental fitness. While the implicit reward did not show a statistically significant correlation with individual $K_\mathrm{D}$ values, binders exhibited a trend toward higher mean implicit reward compared to non-binders (mean difference = 0.0083; Mann–Whitney U test $p=1.28\times10^{-1}$), suggesting that this signal may be informative at the population level.

Finally, we assessed the impact of alignment on enzymatic sequence representations across other Enzyme Commission (EC) classes that were not involved in the alignment objective. For 6,000 different EC labels, 1 sequence was generated using both the pre-trained and the aligned models (after 30 RL iterations), and sequence quality was evaluated using ESM-1v. The spearman correlation between ESM-1v scores of sequences generated by the two models was 0.738, indicating that we did not observe strong evidence of substantial forgetting in this analysis (Fig. \ref{fig:fig5}c). Consistently, the mean ESM-1v score remained largely unchanged (the difference between the mean of the pre-trained model versus the alignment model results to be 0.06). This stability was also observed for enzymatic classes sharing the same EC label up to the second hierarchical level (Fig. \ref{fig:SI_EGFireward}).

In conclusion, we emphasize that implicit reward inherits the limitations of the underlying oracle: alignment may capture CLEAN, or predictor-specific features that are not necessarily biologically causal. Nevertheless, when experimental assays assign fitness at the whole-sequence level (e.g., binding or activity), token-level implicit rewards provide an avenue for amino acid resolution insight, offering suggestive, non definitive, mechanistic hypotheses while maintaining low computational overhead.

\section*{Discussion}
Protein language models (pLMs) are powerful generative priors that efficiently sample from their training distribution, but they lack a principled mechanism to optimize sequences for user-defined properties. Here, we address this limitation by implementing and evaluating ProtRL, an alignment tool that steers autoregressive pLMs towards user-defined properties via interaction with external oracles, including experimental feedback. Across diverse objectives, ProtRL reliably shifted the generation distribution in the desired direction without requiring additional labeled training data, while converging within a small number of iterations.

ProtRL was effective across discrete, bounded, continuous, and multi-objective campaigns while preserving sequence diversity. In a fully experimental, lab-in-the-loop setting, ProtRL improved EGFR-binding affinity across iterative rounds and produced binders that exceed the strongest variant reported in the literature, despite relying on few feedback points. This result highlights ProtRL utility in data-limited regimes, which are typical of protein engineering. More broadly, these findings suggest that, analogous to alignment in NLP, preference- and reward-based optimization offer a practical interface between pLMs and heterogeneous design constraints.

ProtRL offers practical advantages over recursive fine-tuning with synthetic data. RL directly maximizes a reward signal and can learn from both positive and negative outcomes, which are often abundant in protein engineering campaigns. It also does not require curated training corpora: the model improves by learning from its own generations paired with oracle feedback. In addition, we observed that RL applied to a single targeted condition can yield improvements that generalize to related conditions.
This suggests a route to rapidly upgrade existing pLMs with minimal compute and without collecting new labeled data.

Our experiments highlight several implementation considerations that generalize across design tasks. Oracles provide quantitative objectives that enable automated evaluation and steering of generative models, but they also introduce new challenges related to noise, bias, and cost. Single-term rewards can be exploited (e.g., via length inflation), requiring simple auxiliary terms, such as length control and structure-confidence constraints, to stabilize learning. Incorporating plausibility terms (e.g., pLDDT)  prevents optimization toward degenerate sequences, while tracking diversity metrics during training provides an early warning signal for model collapse. Together, these practices make ProtRL readily deployable as a modular component in typical protein-engineering workflows. 

Beyond optimization, ProtRL provides an interpretability handle through the implicit token-level reward. Token-wise rewards highlight positions that drive alignment and can be used to propose residue-level hypotheses for subsequent testing.

As pLM capabilities increase, alignment should also be considered from a safety and security perspective. Dataset leakage and split artifacts can inflate apparent generalization in protein machine learning (ML), motivating leakage-aware evaluation protocols. In addition, results from NLP show that training examples can be retrieved from deployed models under adversarial querying, including from aligned systems. These findings motivate systematic assessment of memorization and retrieval risk for pLM, particularly when the models are trained on proprietary or sensitive sequences, strengthening the need for conservative deployment practices\cite{wangwithout}.

In conclusion, ProtRL provides a generalizable and modular framework for designing high-fitness variants under complex external rewards without imposing explicit constraints on sequence identity or length. By iteratively refining pLMs through interaction with external oracles, ProtRL can optimize objectives such as binding affinity and enzyme activity, even in data-scarce regimes. We envision integrating ProtRL into automated laboratory pipelines where experimental measurements provide direct reward signals, enabling closed-loop agents that propose, test, and improve sequences over time. As in NLP, we expect that continued progress will require not only better oracles and automation, but also systematic alignment and security evaluation as protein generative models become increasingly capable.

\section*{Acknowledgments}
We are most grateful to Adaptyv Bio, for testing our EGFR binders in their protein characterization platform as part of the Adaptyv Bio Protein Design Competition Round 2, and Julian Englert, Daniel Nakhaee-Zadeh Gutierrez, Ron Gill and all the team for assistance. We are grateful to Gerard Boxó, Michael Heinzinger, Maurice Brenner, and Alex Vicente and other members of the Ferruz lab for their insightful discussions. We thank Luis Serrano for discussions on the experimental data of EGFR binders. The authors gratefully acknowledge the scientific support and HPC resources provided by the Erlangen National High-Performance Computing Center (NHR@FAU) of the Friedrich Alexander-Universität Erlangen-Nürnberg (FAU) under the NHR project b114cb (UID 210235). Federal and Bavarian state authorities provide NHR funding. NHR@FAU hardware is partially funded by the German Research Foundation (DFG) – 440719683. We acknowledge the support of the Spanish Ministry of Science and Innovation through the Centro de Excelencia Severo Ochoa (CEX2020-001049-S, MCIN/AEI /10.13039/501100011033), and the Generalitat de Catalunya through the CERCA program. FS is supported by a fellowship from the “la Caixa” Foundation (ID 100010434, fellowship code DFI25-00620F). MG acknowledges funding from project CPP2023-010699 funded by MCIN/ AEI/ 10.13039/ 50110001103312 and by the European Union (FEDER, UE). N.F. acknowledges support from a Ramón y Cajal contract RYC2021-034367-I funded by MCIN/ AEI/ 10.13039/ 501100011033 and by the European Union NextGenerationEU/PRTR. M.A.L. acknowledges support from an FI Fellowship (AGAUR-Catalan Government) co-funded by the European Social Fund (Award 2024 FI-3 00065). We acknowledge support from the ERC-2024-Starting grant ATHENA (GA: 101165231). This project has received funding from the European Union's Horizon Europe under grant agreement No 101120466 (MSCA-DN supporting A.H.). Views and opinions expressed are however, those of the author(s) only and do not necessarily reflect those of the European Union. Neither the European Union nor the granting authority can be held responsible for them.

\section*{Methods}

\subsection*{General pipeline}
The ProtRL framework operates through an iterative generate-score-update cycle that aligns a pre-trained autoregressive protein language model (pLM) with external, user-defined objectives. The general pipeline consists of four primary stages: initialization, sequence generation, oracle scoring, and policy optimization.

\textbf{Initialization.} The process begins with a base autoregressive pLM (e.g., ZymCTRL or ProtGPT3). Depending on the design task, this model may optionally undergo supervised fine-tuning (SFT) on a task-specific sequence dataset (e.g., a known enzyme family or target-specific binders) to narrow the generation distribution toward a relevant prior. This adapted model serves as the reference policy ($\pi_{\text{ref}}$) for subsequent alignment and provides the anchor for Kullback-Leibler (KL) divergence regularization. Usually 200 per iteration. 

\textbf{Sequence Generation.} At each training iteration, the active policy ($\pi_{\theta}$) generates a batch of novel protein sequences. Generation is performed autoregressively, optionally conditioned on contextual prompts such as Enzyme Commission (EC) identifiers or target-specific control tokens.

\textbf{Oracle Scoring.} The generated sequences are independently evaluated by one or more external oracles to assign a scalar reward. Importantly, the reward can be non-differentiable and from broad and different sources (from \textit{in silico} to real world experimental values). For multi-objective tasks, individual oracle scores are aggregated into a composite reward function, frequently incorporating regularizers such as length penalties to prevent reward hacking.

\subsection*{Code availability}
The code used for fine-tuning, training, structure prediction, scoring, and alignment is publicly available at \url{https://github.com/AI4PDLab/ProtRL/} (v2.0.0, branch \texttt{experiments\_v3}).

\subsection*{REINFORCE}
The REINFORCE algorithm optimizes the expected reward by increasing the log-likelihood of sampled actions weighted by their corresponding returns \( r(x) \). The objective can be written as:
\begin{align}
\mathcal{L}_{\mathrm{REINFORCE}} = - \mathbb{E}_{x \sim \pi_\theta} \left[r(x)\,\log \pi_\theta(x) \right]
\end{align}

\subsection*{Direct Preference Optimization (DPO)}
Given a prompt $x$ and a preferred/rejected pair of completions $(y_w, y_l)$ drawn from a preference dataset $\mathcal{D}$, the DPO objective is defined as:
\begin{equation}
\mathcal{L}_{\mathrm{DPO}}(\pi_\theta;\pi_{\mathrm{ref}})
= -\mathbb{E}_{(x,y_w,y_l)\sim \mathcal{D}}
\left[\log \sigma\!\left(\beta\left(\Delta_\theta(x,y_w)-\Delta_\theta(x,y_l)\right)\right)\right],
\label{eq:dpo}
\end{equation}
where $\sigma(\cdot)$ denotes the logistic sigmoid, $\pi_\theta$ is the trainable policy, and $\pi_{\mathrm{ref}}$ is a fixed reference policy. The term
\begin{equation}
\Delta_\theta(x,y)=\log \pi_\theta(y\mid x)-\log \pi_{\mathrm{ref}}(y\mid x)
\label{eq:delta}
\end{equation}
represents the relative log-likelihood between the two policies. The hyperparameter $\beta$ controls the strength of optimization, effectively regulating the deviation of $\pi_\theta$ from $\pi_{\mathrm{ref}}$ \cite{rafailov2023direct}.

\subsection*{Ranked and weighted DPO losses}
\label{RL_expl}
Following Widatalla et al. (2024) \cite{widatalla2024aligning}, we additionally consider listwise extensions of DPO for scenarios in which each prompt \(x\) is associated with an ordered set of \(K\) candidate outputs \(\{y_1, \dots, y_K\}\), ranked from best to worst according to an oracle. Using \(\Delta_\theta(x, y_j)\) as defined in Eq.~\ref{eq:delta}, the ranked loss is given by:

\begin{equation}
\mathcal{L}_{\mathrm{DPO\_ranked}}(\pi_\theta; \pi_{\mathrm{ref}}) = -\mathbb{E}_{(x, \{y_j\}) \sim \mathcal{D}} \left[ \sum_{k=1}^{K} \log \frac{\exp(\beta \, \Delta_\theta(x, y_k))}{\sum_{j=k}^{K} \exp(\beta \, \Delta_\theta(x, y_j))} \right].
\end{equation}

A weighted variant introduces non-negative coefficients \(w_k\) for each rank:
\begin{equation}
\mathcal{L}_{\mathrm{DPO\_weighted}}(\pi_\theta; \pi_{\mathrm{ref}}) = -\mathbb{E}_{(x, \{y_j\}) \sim \mathcal{D}} \left[ \sum_{k=1}^{K} w_k \, \log \frac{\exp(\beta \, \Delta_\theta(x, y_k))}{\sum_{j=k}^{K} \exp(\beta \, \Delta_\theta(x, y_j))} \right].
\end{equation}

In practice, the weights are normalized across each batch using a softmax function, and the loss is implemented using standard PyTorch operations.

\subsection*{Group Relative Policy Optimization (GRPO)}
GRPO is an alternative alignment strategy that extends Proximal Policy Optimization (PPO) by grouping generated outputs into sets of size \( |o| \) and computing advantages relative to the group mean. This formulation reduces gradient variance and improves training stability without requiring a separate value function.

The GRPO objective is defined as:
\begin{equation}
\mathcal{L}_{\mathrm{GRPO}} = -\frac{1}{G} \sum_{i=1}^{G} \frac{1}{|o_i|} \sum_{t=1}^{|o_i|} \min \left( r_{i,t}(\theta) \, \hat{A}_{i,t}, \, \text{clip}(r_{i,t}(\theta), 1-\epsilon, 1+\epsilon) \, \hat{A}_{i,t} \right) + \beta \, D_{\mathrm{KL}} \left( \pi_\theta \| \pi_{\mathrm{ref}} \right),
\end{equation}

where each group \(o_i\) contains trajectories with similar characteristics. The group-relative advantage \(\hat{A}_{i,t}\) is computed as:
\begin{equation}
    \hat{A}_{i,t} = \frac{r_i - \text{mean}_t(r)}{\text{std}_t(r)},
\end{equation}
with \(\text{mean}_t\) and \(\text{std}_t\) evaluated over responses within the same group.

\subsection*{Token-level implicit reward}
We computed the implicit reward as in Rafailov et al.\cite{rafailov2024r}:
\begin{equation}
r_t = \beta\left[\log \pi_\theta(a_t\mid s_t)-\log \pi_{\mathrm{ref}}(a_t\mid s_t)\right],
\label{eq:implicit}
\end{equation}
where $s_t$ is the prefix (state) at position $t$ and $a_t$ is the emitted amino acid (action).

\subsection*{Self-fine-tuning (self-FT)}
Self-fine-tuning iteratively refines ZymCTRL using its own top-scoring generations under a predefined scoring function $f(\mathrm{seq})$. Each iteration consisted of: (i) generating 2{,}000 sequences; (ii) ranking sequences by $f(\mathrm{seq})$; and (iii) fine-tuning on the top 10\% (200 sequences). We repeated this procedure for 30 iterations. All s-FT experiments were run in triplicate.

\subsection*{Model training and sequence generation}
For s-FT, we fine-tuned ZymCTRL using the HuggingFace \texttt{Trainer} (Transformers v4.45.2). We used a training batch size of 4, evaluation batch size of 1, learning rate $8\times10^{-6}$, and trained for 25 epochs per iteration. Models were evaluated and checkpointed every 10 steps; the checkpoint with the lowest evaluation loss seeded the next iteration.

For both ProtRL and s-FT, sequences were sampled with \texttt{top\_k}=9, \texttt{top\_p}=1.0, \texttt{repetition\_penalty}=1.2, and \texttt{do\_sample=True}.

\subsection*{Structural similarity and sequence identity}
Sequences were folded with ESMFold \cite{lin2022language} (Transformers v4.45.2). Structural similarity was assessed by aligning predicted structures to reference Protein Data Bank (PDB) structures using Foldseek \cite{van2024fast} (release 9-427df8a) and reporting TM-score where applicable. Sequence identity to training data was computed by aligning generated sequences against the training set of the corresponding model checkpoint using MMseqs2 \cite{steinegger2017mmseqs2} (release 15-6f452). Diversity was summarized by MMseqs clustering at the identity thresholds reported in the main text.

\subsection*{Functional annotation with CLEAN}
Functional annotation was computed using CLEAN \cite{yu2023enzyme} (v1.0.1). Protein embeddings were first extracted using ESM-1b \cite{rives2021biological}, producing an $L\times d$ embedding matrix for a sequence of length $L$. Embeddings were then projected into CLEAN’s optimized embedding space and scored by cosine similarity to the cluster center corresponding to the target EC label. Where indicated, this cosine similarity served directly as the oracle score.

\subsection*{Likelihood-based metrics: ESM1v and ProteinMPNN}
ESM1v scores were computed as the mean log-probability of the observed amino acid at each position under ESM1v \cite{meier2021language}following established protocols \cite{johnson2025computational}. ProteinMPNN scores were computed using the \texttt{score\_only} mode of ProteinMPNN \cite{dauparas2022robust} (v1.0.1). To ensure consistent maximization objectives, ProteinMPNN negative log-likelihoods were multiplied by $-1$ when used as rewards.

\subsection*{Activity prediction models}
To obtain an experimental proxy reward for $\alpha$-amylase activity, we trained sequence-to-scalar regression models following Schmirler et al.\ (2024) \cite{schmirler2024fine}. We used ESM2 and ESM1v backbones with LoRA adapters. Training data were obtained from the Protein Engineering Tournament dataset \cite{armer2025results}, which provides sequences with Specific Activity Performance Index (SAPI), defined as the ratio of a variant’s specific activity to the reference, with specific activity computed as activity normalized by measured concentration. Entries with missing activity or expression $<0.5$ were removed. Models were trained for 57 epochs with an 80/20 train/validation split, batch size 4, and learning rate $3\times10^{-4}$. The two model predictions were averaged and used as the oracle reward.

\subsection*{\textit{In silico} ProtRL optimization of three binding targets}

\textit{In silico} RL of three different targets was carried out leveraging different \textit{in silico} scores (Methods). Code used to fold, score and align the models is publicly available at \url{https://github.com/AI4PDLab/ProtRL/} (branch experiments\_v3).

\subsubsection*{EGFR Binders - in silico optimization}
ProtGPT3 was fine-tuned on a curated dataset of 500 epidermal growth factor (EGF) sequences (released with the code) using a batch size of 5 for 10 epochs and learning rate of 5e-5. The dataset was constructed by performing a BLAST search using human EGF as a query, followed by clustering at 30\% sequence identity using MMseqs2. Representative sequences from each cluster were used for fine-tuning.

The fine-tuned model was used as the reference model for reinforcement learning. ProtRL (GitHub v2.0.0) was applied using the GRPO algorithm, with the scoring function defined in Table~\ref{Table:SI_loss_func}, a $\beta$ value of 0.5, and a learning rate of $1\times10^{-4}$. EGFR P00533 was used to co-fold with the generated sequences. Generated sequences were folded without templates and MSA using Boltz2, while EGFR was provided with MSA and templates. Scripts, templates and MSA are available in the released code.  

\subsubsection*{Nipah virus glycoprotein G Binders}
ProtGPT3 was fine-tuned using a dataset of 10,000 sequences derived from the wild-type Ephrin-B2 protein, randomly mutated at 8 to 140 positions (dataset released with the code). Fine-tuning was performed with a batch size of 5 for 1 epoch and a learning rate of $1\times10^{-5}$. Nipah virus glycoprotein G (PDB 2VSM) was used to co-fold with the generated sequences. Generated sequences were folded without templates and MSA using Boltz2, while the target was provided with MSA and templates. Scripts, templates and MSA are avaiable in the released code

The resulting model was used as the reference model for ProtRL optimization. ProtRL (v2.0.0) was applied using the REINFORCE algorithm, with the scoring function defined in Table~\ref{Table:SI_loss_func}, a $\beta$ value of 0.1, and a learning rate of $3\times10^{-4}$.

\subsubsection*{CD80 Binders}
ZymCTRL was fine-tuned on a subset of the AlphaFold Atlas comprising sequences with predicted structures satisfying the following criteria: pLDDT score $>80$, non-fragmented sequences, and sequence length between 70 and 100 amino acids. This filtering resulted in 487,081 sequences. The dataset was further clustered at 30\% sequence identity, yielding 143,707 representative sequences used for fine-tuning. Fine tuning was carried out for 10 epochs and a learning rate of $1\times10^{-3}$, with a batch size of 8. 
The resulting model served as the reference model for ProtRL (v1.0.0), applied using the GRPO algorithm with the scoring function defined in Table~\ref{Table:SI_loss_func}, a $\beta$ value of 0.01, and a learning rate of $1\times10^{-6}$. CD80 (PDB 4RWH) was co-fold with the generated sequences. Generated sequences were folded without templates and MSA using Boltz2, while the target was provided with MSA and templates. Scripts, templates and MSA are available in the released code.

\subsection*{EGFR Binder engineering campaign: iterative alignment and optimization}

\subsubsection*{Round 1: SFT and baseline characterization}
A dataset of 500 sequences (available at \href{https://github.com/AI4PDLab/ProtRL/blob/experiments_v3/experiments_v3/EGFR_analysis/clusterResult_EGF_rep_seq.fasta}{clusterResult\_EGF\_rep\_seq.fasta}) was generated via standard BLAST search using wild-type human EGF as the query. Sequences were prepended with the artificial EC label 1.3.3.18. ZymCTRL was fine-tuned on these sequences using a learning rate of $1 \times 10^{-6}$ for 30 epochs with a batch size of 1. A total of 10,000 sequences were generated, and the top six sequences with the lowest perplexity were selected for testing. Training sequences are available in the released code.

\subsubsection*{Round 2: Initial alignment via rDPO}
Model alignment was performed using ProtRL (GitHub releasev1.0.0) with Ranked Direct Preference Optimization (rDPO). The training dataset was derived from the AdaptyvBio competition dataset \cite{cotet2025crowdsourced}, filtered to include candidates with a TM-score greater than 0.4 when compared to the wild-type EGF. To mitigate the imbalance between positive and negative binders, all positive binders were retained, whereas negative binders were subsampled such that their proportion did not exceed a 2:3 ratio relative to the positive set. Sequence scores were computed as described in Table \ref{Table:SI_loss_func}. The training sequences are released together with the code.

Alignment was performed with a $\beta$ value of 0.1, a learning rate of $1\times10^{-7}$, and for 10 and 20 epochs. From the aligned models, 10,000 sequences were generated, and the top 1,000 sequences with the lowest perplexity from each generation round were selected. These sequences were then merged and folded using AlphaFold. The final 20 candidates selected for experimental testing were chosen on the basis of high \textit{in silico} performance, including predicted pLDDT, implicit reward, and additional design-related optimal threshold metrics described in BindCraft \cite{pacesa2024bindcraft}.

\subsubsection*{Rounds 3.1 and 3.2: Calibration and failure analysis}
In Round 3.1, the Round 2 model was used as the starting point for further rDPO; however, this resulted in a significant reduction in hit rate, attributed to likely due to suboptimal hyperparameter tuning. In Round 3.2, we erroneously used the original pre-trained ZymCTRL as the KL-divergence reference model instead of the SFT model. In both rounds, we applied a very low beta of 0.01. These rounds highlighted the necessity of anchoring the alignment to the SFT prior and maintaining strict hyperparameter control.

\subsubsection*{Round 3.3: Final optimization via GRPO and internal signal}
To resolve discrepancies between public datasets and in-house measurements, we executed a final alignment using Group Relative Policy Optimization (GRPO) in ProtRL (GitHub release v2.0.0). The training set was restricted to 37 high-quality measurements from previous rounds, excluding pubblicly avaiable datasets, to ensure a consistent reward signal (alignment dataset shared in the GitHub repository). We utilized a learning rate of $2 \times 10^{-6}$ and an increased $\beta$ from 0.01 to 0.5, and set the correct reference model (Fine tuned ZymCTRL used in round 1) to prevent the drift observed in Round 3.1 and 3.2.
From the resulting aligned model, 10,000 sequences were sampled and ranked by their perplexity and implicit reward:
\begin{equation}
\text{Reward}_{\text{implicit}} = \beta \log \frac{\pi_{\theta}(x)}{\pi_{\text{ref}}(x)}
\end{equation}
with reference model the fine tuned ZymCTRL.
The top 1,000 candidates for each list were then folded and screened via AlphaFold2 and scoring matrices as reported in Methods section \ref{insilicometrics}. The final 22 sequences selected for synthesis demonstrated a significant improvement in hit rate and affinity, yielding four binders that outperformed the literature standard m28.

\subsection*{\textit{In Silico} metrics for binder scoring}
\label{insilicometrics}
Following the folding of generated designs, candidates were comprehensively scored using a multi-parametric suite of \textit{in silico} metrics inspired by the BindCraft pipeline \cite{pacesa2024bindcraft}. These metrics were utilized to assess structural integrity, interface quality, and predicted binding affinity. The scoring criteria can be categorized into four primary domains:

\begin{itemize}
    \item \textbf{Folding Confidence and Sequence Fitness:} To evaluate the self-consistency and structural stability of the binder, we calculated the predicted Local Distance Difference Test ($pLDDT$), Predicted TM-score ($pTM$), and ProteinMPNN ($pMPNN$) scores. Secondary structural elements, such as helicity and total sequence length, were monitored to ensure alignment with the target EGF-like fold.
    
    \item \textbf{Interface Geometry and Physicochemistry:} The physical complementarity between the binder and the EGFR extracellular domain was assessed using $shape\_complementarity$, $interface\_dSASA$ (buried solvent-accessible surface area), and total $binder\_sasa$. We also quantified $uns\_hydrogens$ (unsatisfied buried polar atoms) and surface $hydrophobicity$ to penalize designs with poor desolvation profiles.
    
    \item \textbf{Predicted Binding Quality:} To predict binding efficacy, we leveraged interface-specific metrics including interface $pLDDT$ ($i\_plddt$), interface $pTM$ ($i\_ptm$), and $ipTM$. Additionally, we calculated the $\Delta\Delta G$ of interaction ($delta\_delta\_interaction$) and utilized $pDockQ$ and $pDockQ2$ scores to provide a normalized assessment of docking quality.
    
    \item \textbf{Interface Error and Distance Metrics:} Potential structural deviations were penalized using Predicted Aligned Error (PAE) variants, specifically interface PAE ($i\_pae$), binder-specific PAE ($b\_pae$), and total PAE ($t\_pae$). We further incorporated $ipsae$ (interface positional structural alignment error) and $d0$ distance metrics ($d0res$, $d0chn$, $d0dom$, and $iptm\_d0chn$) to ensure the predicted complex remained within physically realistic coordinate constraints.
\end{itemize}

A combination of these metrics  was used to filter the final candidate pool (detailed in Table \ref{tab:feature_auc} and Fig. \ref{fig:SI_filters}), ensuring that only designs with high structural confidence and favorable interface energetics were selected for experimental validation.

\subsection*{Analysis over binders}
An analysis of the binders was conducted to derive filtering criteria. Specifically, all sequences were folded using AlphaFold, and the previously mentioned metrics were used to determine which best describe the ability to bind and not bind (Fig. \ref{fig:SI_filters}). This analysis was performed with a simple ROC AUC evaluation, which allowed for the identification of thresholds to filter the designs \ref{tab:feature_auc}.

\subsection*{Binder analysis and implicit-reward visualization}
We computed correlation matrices between experimental $K_\mathrm{D}$ and \textit{in silico} metrics using only designs with measured affinity. Sequence identity to known sequences was estimated using BLAST with default settings, reporting the single highest-identity hit per query. Implicit rewards were computed per token using Eq.~\ref{eq:implicit} and used to color structural visualizations.

\subsection*{Affinity measurements}
Binding kinetics ($K_\mathrm{D}$, $k_\mathrm{on}$, $k_\mathrm{off}$) were measured using AdaptyvBio’s automated expression and Bio-Layer Interferometry (BLI) pipeline \cite{cotet2025crowdsourced}. The workflow includes DNA optimization for expression, cell-free protein expression, and BLI measurements over a range of EGFR concentrations. Replicates correspond to independently expressed samples. All values, expressions, and sequence types are reported in the supplementary material (see Table \ref{table:sequences_EGFR}).
\subsection*{Compute}
Unless otherwise stated, training was performed on an NVIDIA H100 GPU. ProtRL experiments typically required a few GPU hours to one day depending on the task and number of iterations, while s-FT runs required multiple days to complete triplicate 30-iteration campaigns. All GRPO hyperparameters not specified in the main text are provided in Table \ref{Table:GRPO_hyperparameters}.

\bibliography{bib}

\newpage
\clearpage
\setcounter{page}{1}

\begin{center}
{\Large \textbf{\textit{Supporting Information for:}}}\\[1.5em]

{\LARGE \textbf{Reinforcement Learning Guides Generative Protein Language Models}}\\[1.5em]

Filippo Stocco$^{1,2}$, Maria Artigues-Lleixà$^{2,1}$, Andrea Hunklinger$^{1,3}$, Michele Garibbo$^{1,2}$, Talal Widatalla$^{4,5}$, Marc Güell$^{2,6}$, and Noelia Ferruz$^{*,1,2}$\\[1em]

$^1$Centre for Genomic Regulation, the Barcelona Institute of Science and Technology, Barcelona, Spain\\
$^2$Universitat Pompeu Fabra, Barcelona, Spain\\
$^3$Universitat de Barcelona, Facultat de Farmàcia i Ciències de l’Alimentació, Barcelona, Spain\\
$^4$Stanford University\\
$^5$Arc Institute\\
$^6$Institució Catalana de Recerca i Estudis Avançats (ICREA)\\[1em]

$^*$Correspondence: noelia.ferruz@crg.eu
\end{center}

\vspace{2em}

\setcounter{figure}{0}
\setcounter{table}{0}
\renewcommand{\thefigure}{S\arabic{figure}}
\renewcommand{\thetable}{S\arabic{table}}

\begin{figure}[h]
  \centering
  \includegraphics[width=\textwidth,page=3, trim=0cm 0cm 0cm 0cm, clip]{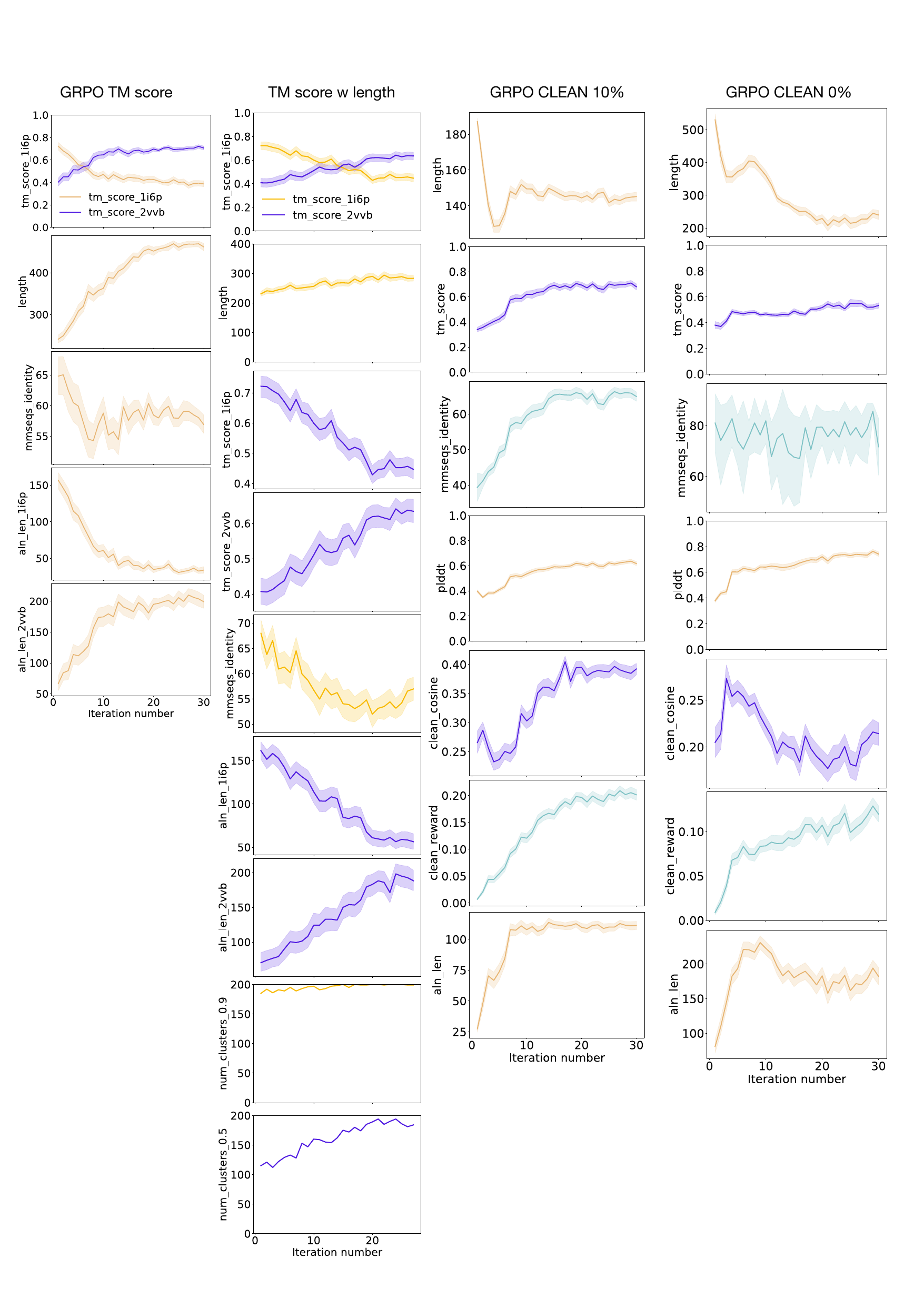}
  \caption{\textbf{Independent ProtRL optimization campaigns targeting distinct protein engineering objectives.}}
  \label{fig:SI_tmscoreGRPO}
\end{figure}
\begin{landscape}
\small
\begin{longtable}{p{1.8cm} p{4.2cm} p{12.2cm}}
\caption{Different scoring functions used in this study. Where $g$ represents the generated sequence. Experimental results are also available at \url{https://proteinbase.com/crg-ferruz-lab}.}
\label{Table:SI_loss_func}\\
\toprule
EC number & Objective & Scoring function \\
\midrule
\endfirsthead

\toprule
EC number & Objective & Scoring function \\
\midrule
\endhead
          
4.2.1.1 & From beta to alpha CA &
$$
\frac{TM_{2vvb} + \frac{align_{2vvb}}{100} - \left| \frac{length_g - length_{2vvb}}{100} \right|}{3}
$$
\\

4.6.1.18 & From 10\% to 100\% CLEAN prediction &
$$
\mathrm{CosSim}(emb_g, emb_{11ba}) \cdot \mathrm{length}_{\mathrm{rew}} \cdot \mathrm{pLDDT},
\qquad
\mathrm{length}_{\mathrm{rew}}
=
\exp\!\left(
-\frac{\left(\frac{align_{11ba}}{length_g}-1\right)^2}{0.5^2}
\right)
$$
\\

2.7.11.5 & From 0\% to 100\% &
$$
\mathrm{CosSim}(emb_g, emb_{3eps}) \cdot \mathrm{length}_{\mathrm{rew}} \cdot \mathrm{pLDDT},
\qquad
\mathrm{length}_{\mathrm{rew}}
=
\exp\!\left(
-\frac{\left(\frac{align_{3eps}}{length_g}-1\right)^2}{0.5^2}
\right)
$$
\\

5.4.99.5 & Increase ProteinMPNN &
$$
-1.0 \cdot \mathrm{ProteinMPNN}_{\mathrm{score}}
+ \mathrm{TM}_{1com}
+ \frac{\mathrm{align}_{1com}}{100}
$$
\\

4.2.1.1 & Increase ESM1v &
ESM1v score
\\

3.2.1.1 & Increase activity (SAPI) &
Predicted activity
\\

- & EGFR binders &
$
\begin{aligned}
\mathrm{score} ={}&\frac{1}{2}
\Big(
- 0,5 \;{length\_norm}
+10\,\mathrm{ipsae}
\Big) 
\end{aligned}
$
\\

- & Nipah virus' glycoprotein G. &
$
\begin{aligned}
\mathrm{score}
={}&
\frac{1}{9}
\Big(
-0.10\,\mathrm{length\_norm}
-\mathrm{b\_pae}
-\mathrm{i\_pae}
+0.30\,\mathrm{shape\_comp} \\
&\quad
+0.70\,\mathrm{int\_dSASA}
-\mathrm{d\_rmsd}
+10\,\mathrm{ipsae}
+\mathrm{lis}
+\mathrm{iptm\_d0chn}
-\mathrm{t\_pae}
\Big)
\end{aligned}
$
\\

- & CD80 binders &
$
\begin{aligned}
\mathrm{score}
={}&
\frac{2\,L_{\mathrm{norm}}}{15}
\Big(
0.80\,\mathrm{plddt}
+0.80\,i_{\mathrm{pTM}}
-0.80\,\mathrm{pAE}_b
-0.80\,\mathrm{pAE}_{bt}
-0.80\,i_{\mathrm{pae}} \\
&\quad
-0.50\,\mathrm{pMPNN}
+0.30\,\mathrm{shape\_comp}
-0.40\,\mathrm{hydrophobicity}
-0.30\,\mathrm{int\_dSASA} \\
&\quad
-0.70\,\mathrm{uns\_hydr}
+0.50\,\mathrm{contact\_probs}
-0.30\,d_{\mathrm{rmsd}}
-0.50\,\mathrm{binder\_sasa} \\
&\quad
+0.60\,\Delta\Delta_{\mathrm{inter}}
-0.50\,\mathrm{helicity}
\Big)
\end{aligned}
$
\\

1.3.3.18  & EGFR binders (in vitro) &
$
\mathrm{score}
=
100\left(
\exp\!\bigl(\max(-\log_{10}(kd)-5,\ 0)\bigr)-1
\right)
$
\\

\bottomrule
\end{longtable}
\end{landscape}

\begin{figure}[h]
  \centering
  \includegraphics[
    width=\textwidth,
    page=6,
    trim=0cm 30cm 0cm 0cm,
    clip
  ]{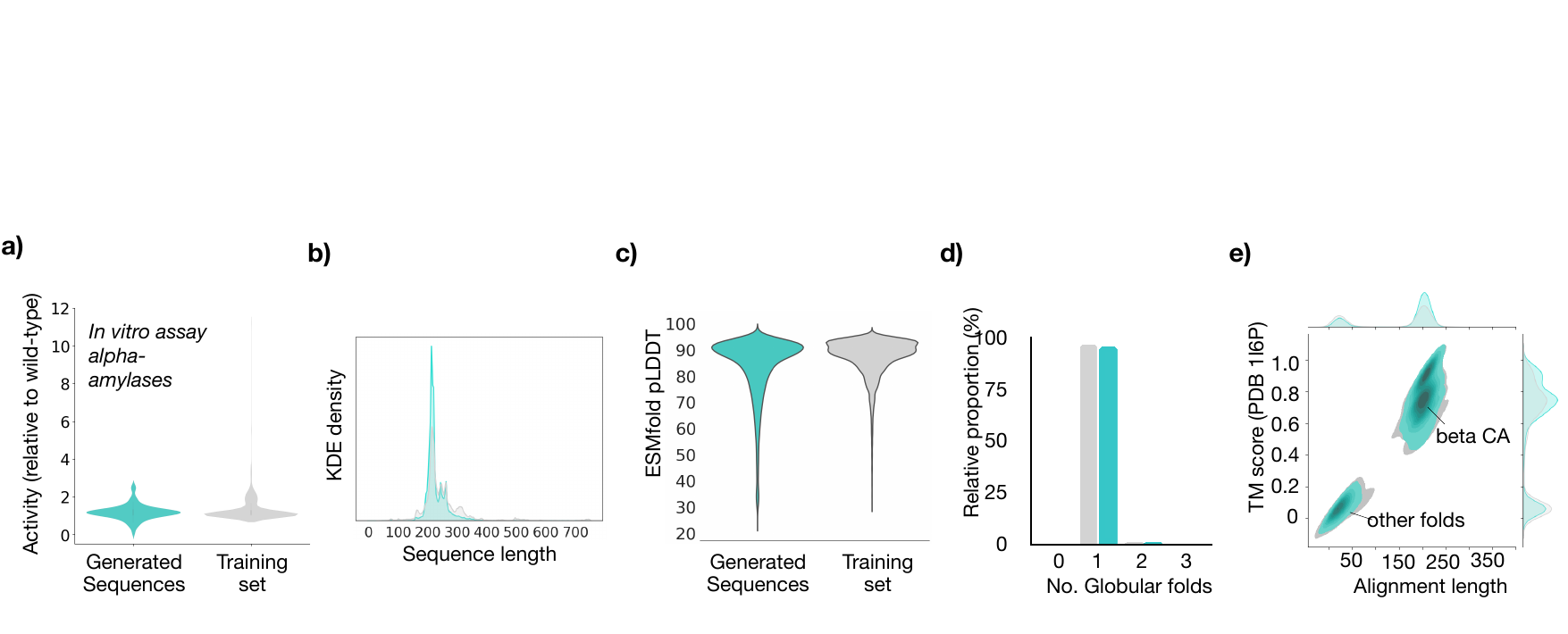}
  
  \caption{\textbf{Distributions of key properties for generated sequences compared with natural proteins.} (a) Experimental activity of generated and natural $\alpha$-amylases. (b) Sequence length distributions. (c) Predicted pLDDT values. (d) Number of structural domains. (e) Topological classes. Across properties, generated sequences mirror the dominant modes of the training distribution, highlighting limited intrinsic control over low-probability regions.}
  \label{fig:SI_alphaamylase}
\end{figure}

\begin{figure}[h]
  \centering
  \includegraphics[width=\textwidth]{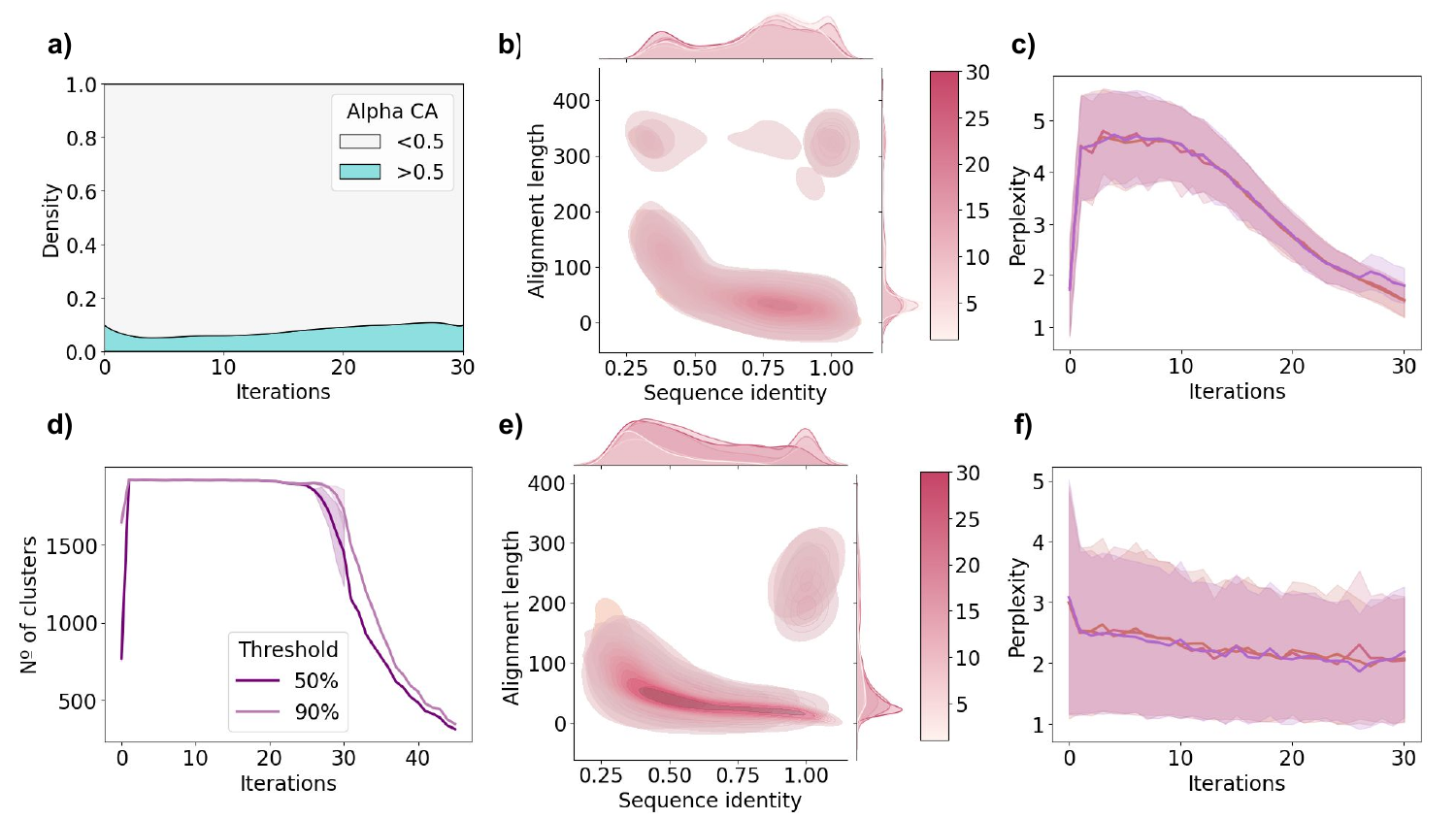}
  \caption{\textbf{Self-FT-guided topology and enzyme functional annotation generation.} (a) Proportion of generated sequences with more or less than 0.5 TM-score when superimposed with PDB 2VVB with TM-score as a reward function. (b) Distribution of sequence similarity of synthetic sequences generated against the 200 sequences set used to finetune the model the previous iteration for the TM-score reward function. (c) Perplexity of the generated sequences with the TM-score reward function across 30 iterations. (d) Progression of the number of clusters of sequences with TM-score reward function until 45 iterations. (e) Distribution of sequence similarity of synthetic sequences generated against the 200 sequences set used to finetune the model the previous iteration got the CLEAN reward function. (f) Perplexity of the generated sequences with the CLEAN reward function across 30 iterations.}
  \label{fig:SI_SFT}
\end{figure}

\begin{figure}[h]
  \centering
  \includegraphics[width=\textwidth,page=1, trim=0cm 4cm 0cm 0cm, clip]{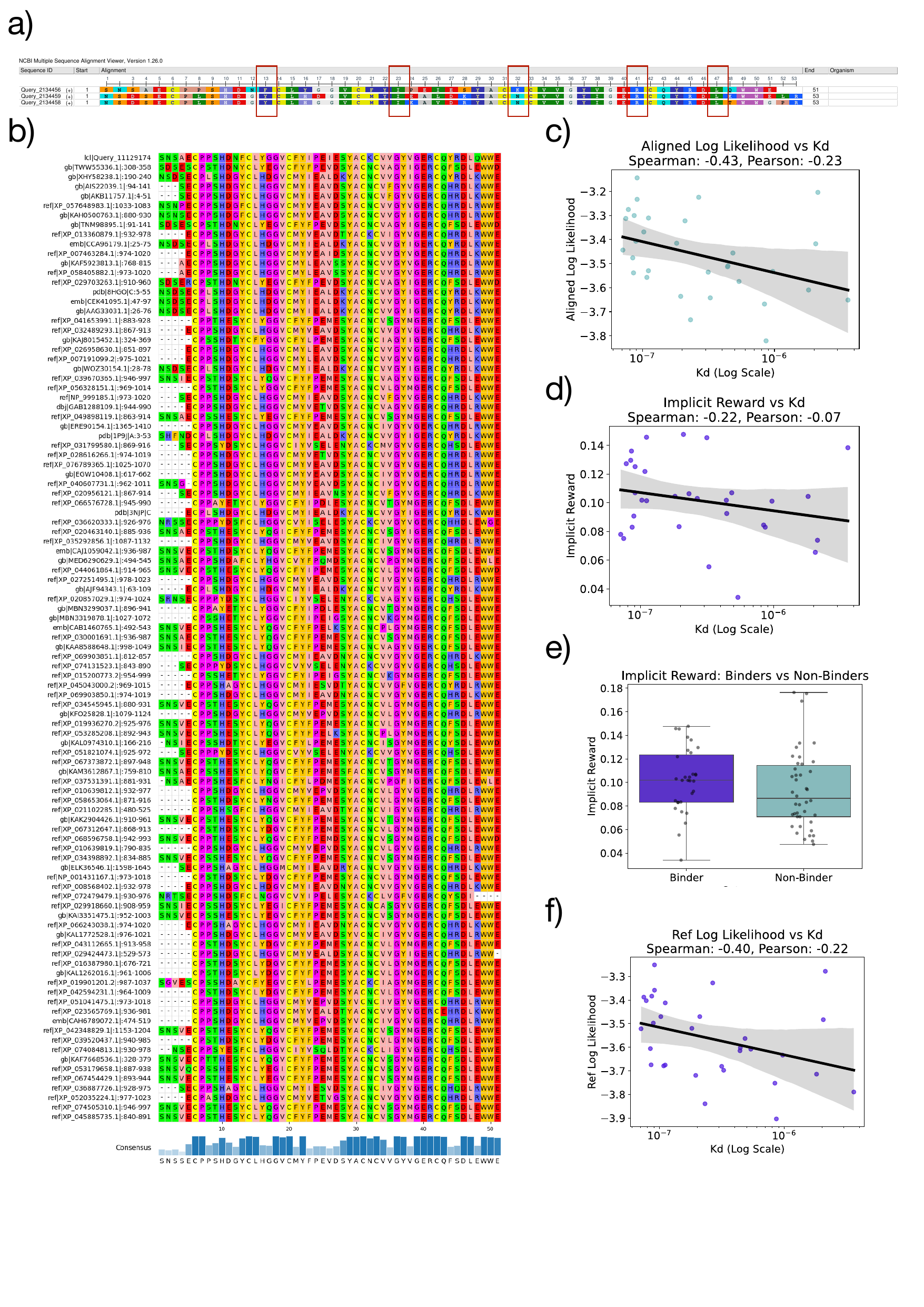}
  \caption{\textbf{Sequence analysis of the top-performing binder.} (a) Pairwise alignment of the highest-affinity designed sequence, the wild-type (WT) reference (second row) and m28 (third row). Red boxes indicate regions previously reported to contribute to binding. (b) Multiple sequence alignment (MSA) of the top 100 homologous sequences retrieved using BLAST. (c) Correlation between sequence log-likelihood and experimental dissociation constant ($K_\mathrm{D}$). (d) Correlation between the implicit reward assigned by the aligned model and experimental $K_\mathrm{D}$ values. (e) Comparison of implicit reward distributions between binders and non-binders.}
  \label{fig:SI_MSA_EGF}
\end{figure}

\begin{figure}[h]
  \centering
  \includegraphics[width=\textwidth,page=2, trim=0cm 0cm 0cm 0cm, clip]{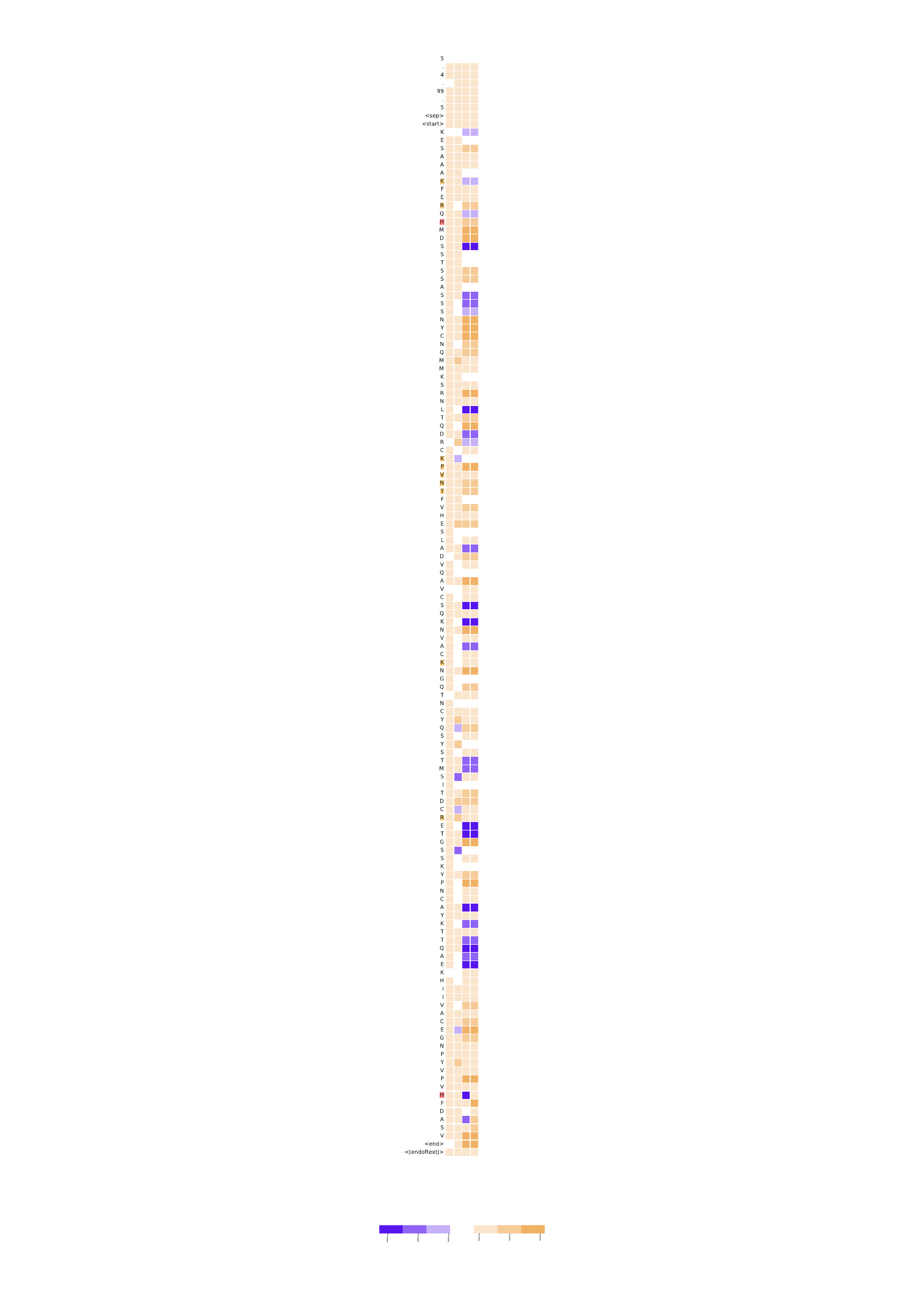}
  \caption{\textbf{Extraction of relevant features at the token level.} Wild-type sequence for Pancreatic Ribonuclease (EC 4.6.1.18, UniProt: P67926). Experiments analyzing single-token variations and their effects compared to the wild-type (WT). Point variation (white asterisks) induce localized changes in the reward, whereas alterations of tokens at the EC label result in a complete drop in the reward across the sequence to zero.}
  \label{fig:SI_mutations_i_reward}
\end{figure}

\begin{figure}[h]
    \centering
    \includegraphics[width=\textwidth,page=4, trim=0cm 8cm 0cm 0cm, clip]{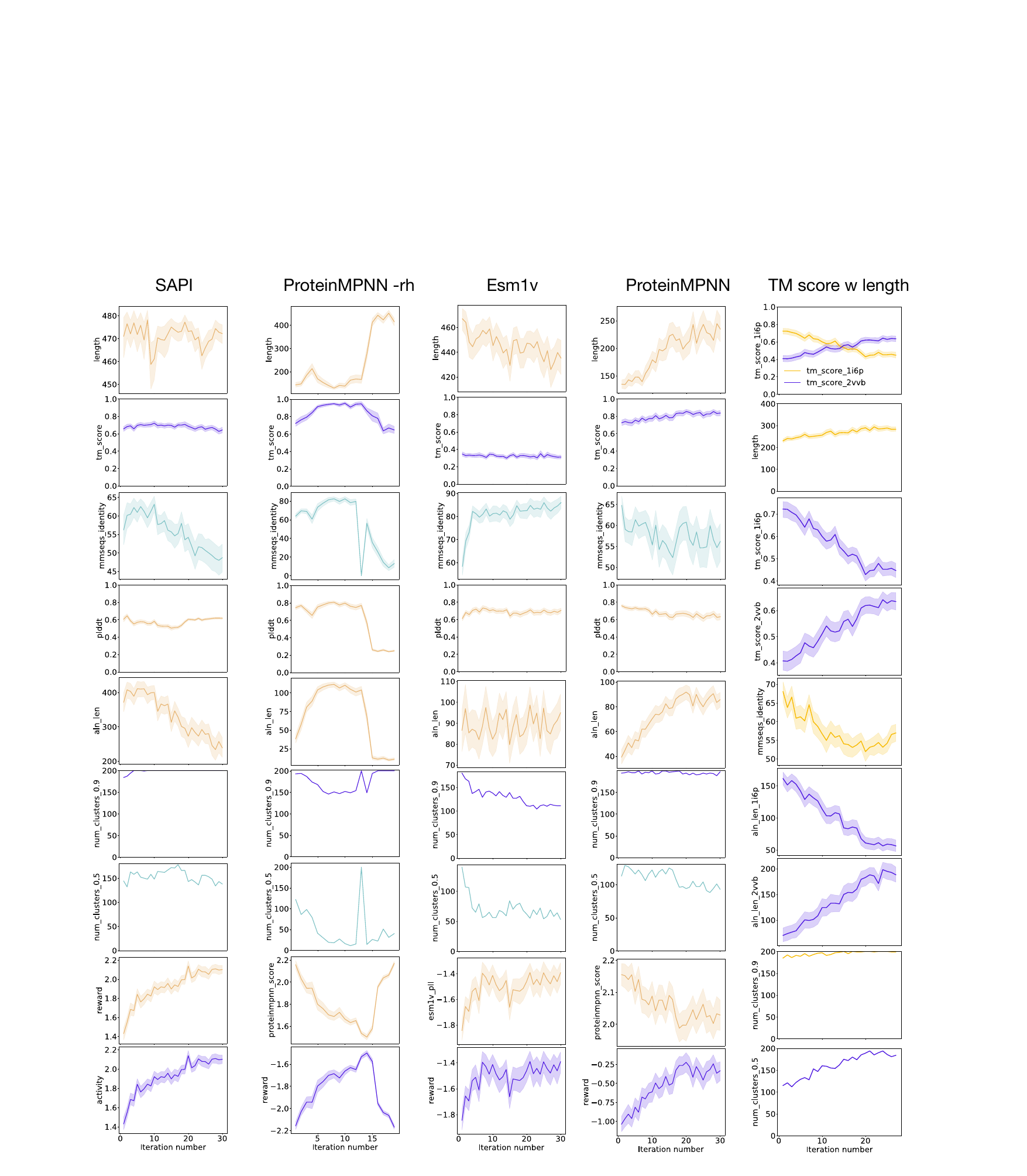}
    \caption{\textbf{Independent ProtRL optimization campaigns targeting distinct protein engineering objectives.} The second column (\textit{ProteinMPNN -rh }) indicates the "reward-hacking" example where the length was not clipped.}
    \label{fig:SI_SAPIGRPO}
\end{figure}

\begin{figure}[h]
    \centering
    \includegraphics[width=\textwidth,page=5, trim=0cm 20cm 0cm 0cm, clip]{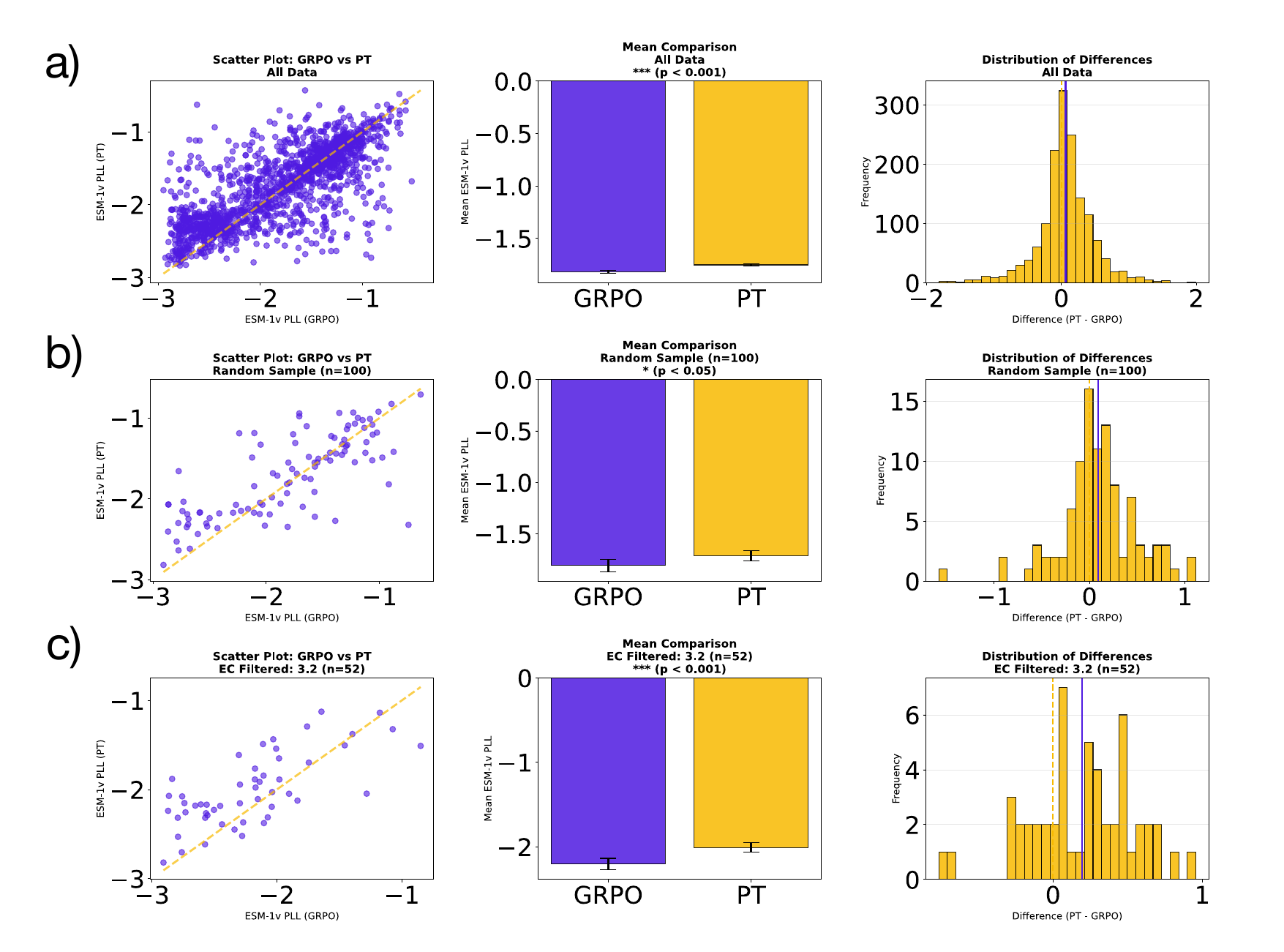}
    \caption{\textbf{Effect of reinforcement learning alignment across EC labels.} ZymCTRL was aligned via reinforcement learning to increase the ESM-1v log-likelihood for a specific Enzyme Commission (EC) label. We evaluated how this alignment influenced score distributions across other EC labels. (a) ESM-1v scores for 10,000 distinct EC labels. (b) A random subset of 100 EC labels. (c) A subset of EC labels sharing the first three hierarchical digits with the target EC label used during reinforcement learning. Each dot represents a sequence generated under the same EC conditioning label by either the pre-trained model (PT) or the aligned model (GRPO), and subsequently scored using ESM-1v log-likelihood. For each group, both the score distribution and the mean value (bar plot) are shown. Alignment increases scores for the target functional class without showing strong evidence of substantial forgetting  across unrelated EC labels.}
    \label{fig:SI_EGFireward}
\end{figure}

\begin{figure}
    \centering
    \includegraphics[width=0.5\linewidth]{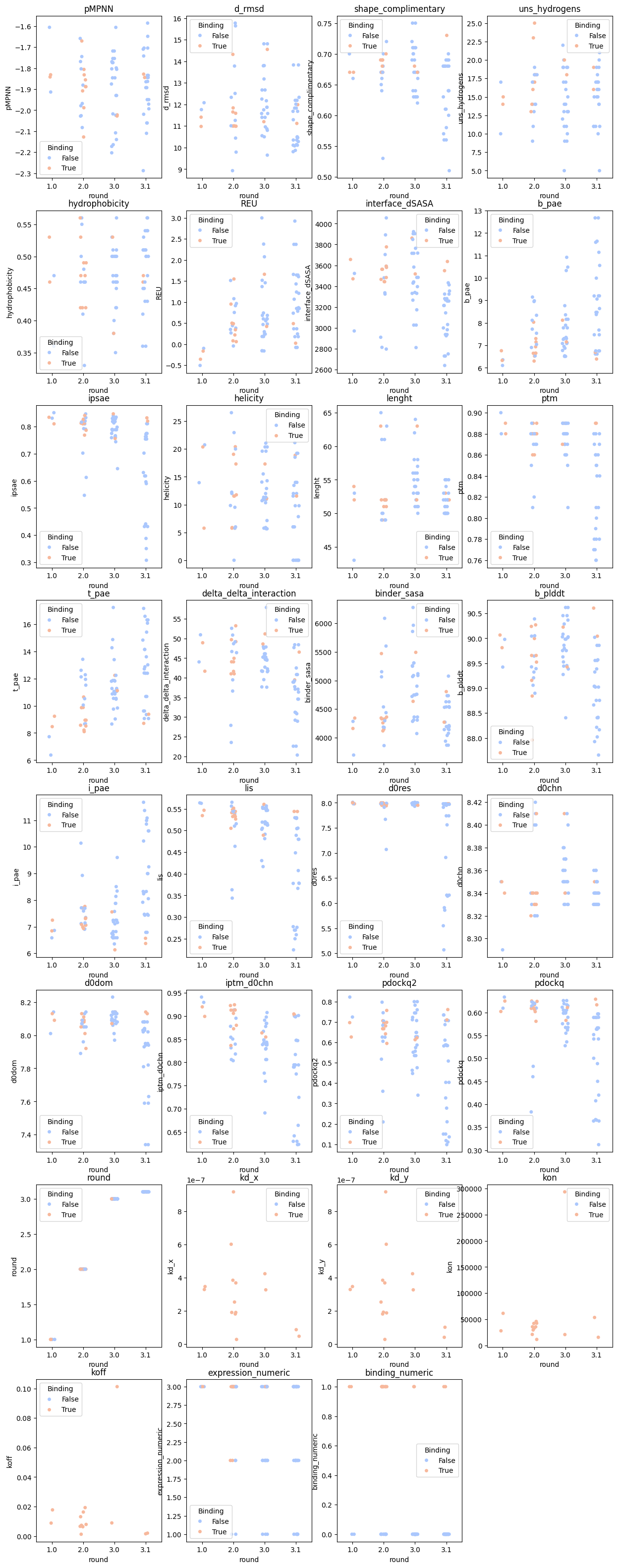}
    \caption{\textbf{Analysis across multiple \textit{in silico} scoring metrics indicates that a subset of these features can effectively discriminate between potential binders and non-binders.} Different rounds of designs were experimentally tested and analyzed. In red the binders and in blue the non binders.}
    \label{fig:SI_filters}
\end{figure}

\begin{table}[h]
\centering
\caption{\textbf{Performance of \textit{in silico} features for discriminating binders from non-binders. }
AUC values were computed from ROC analysis. Direction indicates whether higher or lower values are associated with binding.}
\begin{tabular}{lccc}
\toprule
Feature & AUC & Optimal threshold & Direction for binding \\
\midrule
ipsae & 0.777 & 0.80406 & Higher \\
i\_pae & 0.755 & 7.32175 & Lower \\
lis & 0.747 & 0.5297 & Higher \\
iptm\_d0chn & 0.735 & 0.85487 & Higher \\
ptm & 0.726 & 0.88 & Higher \\
b\_pae & 0.723 & 7.30078 & Lower \\
t\_pae & 0.690 & 9.97736 & Lower \\
b\_plddt & 0.688 & 89.91084 & Higher \\
pdockq2 & 0.663 & 0.5951 & Higher \\
REU & 0.654 & -0.26235 & Lower \\
delta\_delta\_interaction & 0.650 & 40.66516 & Higher \\
pdockq & 0.640 & 0.6019 & Higher \\
shape\_complimentary & 0.640 & 0.67 & Higher \\
d0dom & 0.631 & 8.07 & Higher \\
interface\_dSASA & 0.611 & 3205.82 & Higher \\
\bottomrule
\end{tabular}
\label{tab:feature_auc}
\end{table}

\begin{landscape}
\footnotesize
\ttfamily
\begin{longtable}{lrrrlllr}
\caption{Experimental values for all the sequences tested at each rounds. Control sequences are m28 and m123. WT sequence is wild-type human EGFR.} \label{table:sequences_EGFR} \\
\toprule
round & koff & kon & kd & sequence & expression & confidence & replicate \\
\midrule
\endfirsthead
\caption[]{Experimental values for all the sequences tested at each rounds. Control sequences are m28 and m123. WT sequence is wild-type human EGFR.} \\
\toprule
round & koff & kon & kd & sequence & expression & confidence & replicate \\
\midrule
\endhead
\midrule
\multicolumn{8}{r}{Continued on next page} \\
\midrule
\endfoot
\bottomrule
\endlastfoot
3.3 & $3.72\times 10^{-3}$ & $1.55\times 10^{5}$ & $2.40\times 10^{-8}$ & GNSECPTSHADFCLNNGICIYIEKVDRYACKCVVGYIGERCQFSDLEWWE & medium & high & 3 \\
3.1 & $1.33\times 10^{-2}$ & $1.81\times 10^{4}$ & $7.35\times 10^{-7}$ & NPSDSVAGSTRPGLAEPCPMEYYSPYCLHGECKFIKELSSYACNCVKGYLGERCEFSDLEWWE & high & low & 3 \\
WT & $3.13\times 10^{-3}$ & $6.61\times 10^{4}$ & $4.73\times 10^{-8}$ & NSDSECPLSHDGYCLHDGVCMYIEALDKYACNCVVGYIGERCQYRDLKWWELR & high & high & 2 \\
m28 & $9.07\times 10^{-4}$ & $7.90\times 10^{4}$ & $1.15\times 10^{-8}$ & NSDSECPLSHDGYCLHGGVCMYIKAVDRYACNCVVGYIGERCQYRDLTWWGPR & high & high & 3 \\
3.3 & $8.93\times 10^{-3}$ & $9.73\times 10^{4}$ & $9.17\times 10^{-8}$ & NSNADCPTSHDGYCLHDAVCFYFPEMDSYACNCVVGYVGERCQYKDLKWWDLQ & high & high & 3 \\
3.3 & $4.75\times 10^{-3}$ & $1.12\times 10^{5}$ & $4.23\times 10^{-8}$ & NSNSECPPSHQNFCLHDGVCFYFPEMESFACKCVVGYVGERCQYSDLNWWE & high & high & 3 \\
2 & $2.53\times 10^{-1}$ & $1.76\times 10^{5}$ & $1.43\times 10^{-6}$ & NSVERCPSSHDTYCLYGGVCFYYPETDSFACQCVNGYLGERCQYADLEWWE & high & low & 3 \\
2 & - & - & - & NSVETCPSTYCLHGGICMFIEAVDRYACKCVVGYVGERCQYRDLRWWS & medium & high & 3 \\
2 & $2.03\times 10^{-2}$ & $1.25\times 10^{4}$ & $1.62\times 10^{-6}$ & NSVETCPSTYCLHGGICMFIEAVDRYACKCVVGYVGERCQYRDLRWWSL & medium & high & 2 \\
2 & $1.40\times 10^{-2}$ & $7.21\times 10^{4}$ & $1.94\times 10^{-7}$ & NSVSDCPASHETFCLNGGTCIYFQLLQEYACRCVKGYMGDRCQFSDLEWWE & high & high & 3 \\
m123 & $5.88\times 10^{-3}$ & $7.27\times 10^{4}$ & $8.09\times 10^{-8}$ & NSYSECPPSYDGYCLHDGVCRYIEALDSYACNCVVGYAGERCQYRDLRWWGRR & high & high & 3 \\
1 & $7.33\times 10^{-3}$ & $1.66\times 10^{5}$ & $4.41\times 10^{-8}$ & PDTIASCPSSHESYCLYQGVCFYFPEMESYACNCVSGYMGERCQFSDLQWWELQ & high & high & 3 \\
1 & $5.50\times 10^{-3}$ & $1.04\times 10^{5}$ & $5.31\times 10^{-8}$ & PENYKSCPPSHDGYCLHGGVCMYISALQAYACNCIQGFVGERCQHSDLEWWE & high & high & 3 \\
2 & $1.31\times 10^{-2}$ & $6.57\times 10^{4}$ & $1.99\times 10^{-7}$ & PNSTERCPPSHDSYCLNGGVCIFFSVLQEFACKCLQGYMGERCQHSDLEWWE & high & high & 3 \\
3.3 & $6.21\times 10^{-3}$ & $8.56\times 10^{4}$ & $7.25\times 10^{-8}$ & PNSVESCPSSHDTYCLYDGACRFIPEMEDYSCKCVAGYIGERCQFSDLEWLELQ & high & high & 3 \\
2 & $1.45\times 10^{-2}$ & $5.60\times 10^{4}$ & $2.59\times 10^{-7}$ & PNTVQSCPPSYDSFCLHGGVCIYIPGLEAYACRCVEGYVGERCQFSDLEWWE & medium & high & 3 \\
1 & $7.99\times 10^{-3}$ & $1.32\times 10^{5}$ & $6.03\times 10^{-8}$ & PSTVESCPLSHEGYCLYEGVCFYFPEMESYACNCVSGYMGERCQFSDLEWLELQ & medium & high & 3 \\
3.2 & - & - & - & PTDSSTVGSCPSPAENYCLNGGICFYFPNIQDHACHCASGYIGQRCQYNDLQWWE & medium & - & 3 \\
2 & $1.17\times 10^{-2}$ & $3.39\times 10^{4}$ & $3.45\times 10^{-7}$ & SDCPEAHEGFCLNGGTCIFYEVVKEYACKCLVGFIGDRCQHRDLRWWDL & medium & low & 2 \\
3.3 & $2.98\times 10^{-3}$ & $1.13\times 10^{5}$ & $2.62\times 10^{-8}$ & SNAIERCPPSHDAYCLYGGVCFYIPEMESFACKCVKGYVGERCQFNDLEWWE & high & high & 3 \\
3.3 & - & - & - & SNAVEKCPPSHDDFCLYGGVCSYIVDLQTFACRCVVGYIGDRCRYKDLKWWE & high & - & 3 \\
3.3 & $1.32\times 10^{-3}$ & $1.35\times 10^{5}$ & $9.83\times 10^{-9}$ & SNAVERCPSSHQTYCLYDGVCFYIPEMESFACKCVKGYMGERCQFSDLEWWE & high & high & 2 \\
3.3 & $7.98\times 10^{-4}$ & $1.45\times 10^{5}$ & $5.50\times 10^{-9}$ & SNSAECPPSHDNFCLYGGVCFYIPEIESYACKCVVGYVGERCQYRDLQWWE & high & high & 1 \\
3.2 & $5.77\times 10^{-2}$ & $2.79\times 10^{4}$ & $2.07\times 10^{-6}$ & SNSVDSCPSSHETYCLNGGICLYYDAVKDFACKCVVGFIGDRCQHRDLRWMK & high & high & 2 \\
3.3 & $1.05\times 10^{-2}$ & $3.09\times 10^{4}$ & $3.40\times 10^{-7}$ & SNSVESCPASYGSFCLHDGVCTHYAILQEYACKCVVGFVGERCQYRDLRWWE & medium & low & 3 \\
3.3 & $4.04\times 10^{-3}$ & $9.44\times 10^{4}$ & $4.28\times 10^{-8}$ & SNSVESCPSSHDAYCLYGGVCFYIPEMESFACKCVTGYIGERCQFSDLEWWE & high & high & 3 \\
3.3 & $7.72\times 10^{-3}$ & $1.38\times 10^{5}$ & $5.58\times 10^{-8}$ & SNSVESCPSSHDAYCLYNGICFYFPEMESFACNCITGYMGERCQFNDLKWWE & high & low & 3 \\
3.3 & $6.53\times 10^{-3}$ & $1.66\times 10^{5}$ & $3.94\times 10^{-8}$ & SNSVESCPSSHDTYCLYNGICFYIPEMESFACNCLTGYLGERCQFSDLEWWE & high & high & 3 \\
3.3 & $8.23\times 10^{-3}$ & $1.35\times 10^{5}$ & $6.10\times 10^{-8}$ & SNSVESCPSSHDTYCLYSGICFYFPEMDSFACNCVNGYVGERCQFSDLEWWE & high & high & 3 \\
3.3 & $1.85\times 10^{-3}$ & $9.14\times 10^{4}$ & $2.03\times 10^{-8}$ & SNSVESCPSTHDNYCLYKGICFYIPELESYACNCVTGYMGERCQFSDLEWWE & high & high & 3 \\
2 & $1.61\times 10^{-2}$ & $1.38\times 10^{5}$ & $1.17\times 10^{-7}$ & SNSVESCPSTHDSYCLNGGSCVFFEIVRQFACKCLSGYMGERCQFSDLEWWD & high & high & 3 \\
3.3 & $9.06\times 10^{-3}$ & $9.07\times 10^{4}$ & $9.99\times 10^{-8}$ & SNTTERCPSSHDAYCLYEGVCFYIEAIDQYACKCMLGFVGERCQYSDLEWWE & high & high & 3 \\
2 & $1.44\times 10^{-3}$ & $1.17\times 10^{5}$ & $1.23\times 10^{-8}$ & SNTVQRCPSSHDAYCLYGGVCFYIPEMESFACKCVKGFVGERCQFSDLEWWE & high & high & 3 \\
3.3 & $4.87\times 10^{-3}$ & $1.35\times 10^{5}$ & $3.61\times 10^{-8}$ & SSTVQSCPSSHDTYCLYGGVCFYIPEMESFACNCLKGYMGERCQFSDLQWWE & medium & high & 3 \\
3.3 & $6.37\times 10^{-3}$ & $2.11\times 10^{5}$ & $3.03\times 10^{-8}$ & STPLNDCPPSHDAYCLYEGICFYFPEMESFACNCVKGYMGERCQFSDLKWWELQ & medium & low & 3 \\
3.1 & $3.09\times 10^{-2}$ & $1.18\times 10^{5}$ & $2.61\times 10^{-7}$ & TAEEDKCPPSHESYCLFHGVCRFIEELNLYSCNCIKGFAGERCQFSDLEWWD & high & high & 2 \\
2 & $1.09\times 10^{-2}$ & $9.79\times 10^{4}$ & $1.11\times 10^{-7}$ & TAPESNYCLNETVIDCPTSHDTFCLYDGTCMFIPEMEKYACNCLAGFMGDRCQFSDLQWWEMQ & high & high & 3 \\
3.2 & - & - & - & AGTPCPLDSNYCLNGGVCLYFSIVQEYACQCVVGFVGSRCEHKDVQWWDLR & medium & - & 3 \\
2 & - & - & - & ANVVLDSSPLETCPSTHESYCLYGGKCRFIIQEEMPACRCMKGYDGARCE & high & - & 3 \\
2 & - & - & - & APDADVRLGAGNGTGSECPASYCFKNGTCIFYEPANEFACQCAQGYMGERCQFSDLEWWSR & low & - & 3 \\
3.1 & - & - & - & ASGERRRLDCPPSYAKWYCLNGATCSYIPELQRYGCSCVKGFMGPRCEFSDLSWWEL & medium & - & 3 \\
2 & - & - & - & ASGNPCPMPESYCLHGGVCFYFPDMQSFACRCVSGYIGERCQFSDLEWWA & high & - & 3 \\
3.2 & - & - & - & AVDSCPSTHENYCLNGAICFHVEIVNQYACNCLRGYVGARCEFKDFNWWE & medium & - & 3 \\
3.2 & - & - & - & CSDARTNFGEPCPMSYCLNGGICIYYQVVGKFVCRCQEGFTGRRCESKDV & high & - & 3 \\
3.2 & - & - & - & DSSSLVDCPPSYCLNGGICFYLKPLQEYACKHNNVCKCMPGFTGSRCESKEV & high & - & 3 \\
3.1 & - & - & - & DVKPLASCPEEFRNYCVHGTCRFLKKIGEPACLCKAGYLGQRCEYSDLSWEE & high & - & 3 \\
3.1 & - & - & - & DVNATCPPSYSNYCLHGTCRLLQDLKAPSCECDLGFYGSRCETLELISLE & high & - & 3 \\
3.1 & - & - & - & DVNLGSEERCPLTSYCLHGDCVYYPEMEKYACKCIKGYLGQRCEHSDLEWFE & high & - & 3 \\
3.1 & - & - & - & EKTDINVALGSRSYLNSCPEELTHFCIHGNCRYMEKIQEAVCRCSEGYLGTRCEYKDL & high & - & 3 \\
3.1 & - & - & - & GAVRSTPSLDNLKCPEEFRDYCLHGGECRFIQEQDTLSCKCASGYTGLRCEHSELLTY & none & - & 3 \\
3.1 & - & - & - & GEPALSTPSIDCPAKFSNYCVHGQCLFLEEKNKPACLCMQGFEGERCEFSDLEWWE & none & - & 3 \\
2 & - & - & - & GNSAETSRCPASHDTYCLNGGTCTFYEAVGELICQCAEGYKGKRCENKDV & high & - & 3 \\
3.2 & - & - & - & GPSALSHPCTAEQDGFCLNGGVCIYFEAVGQSACKCAEGYRGERCQNKDV & low & - & 3 \\
3.2 & - & - & - & GSPCPMADSFCLNESTAFCLNGGVCFYIQKIQEVSCSCAAGFIGKRCQHESV & medium & - & 3 \\
3.1 & - & - & - & GVAEPLRGRTCPSPEDFCLNGGSCIYFEPLKEYACQCVEGYEGARCDSKEL & none & - & 3 \\
3.1 & - & - & - & MDCPALEGVDGARLSAKCPANFCLHQGTCIYFADLEKYACQCVEGYQGSRCEHEELEWWELR & medium & - & 3 \\
3.1 & - & - & - & MDLEQRTARSCPPSYCFHDGKCVYLPDLREYVCNCMEGYLGARCEFNDLEWWELK & high & - & 3 \\
2 & - & - & - & MSTPLATLAAASTAGRPCPTEFNPYCLHDGVCIYIPELQKYACDCVRGYMGERCQFSDLQWWELR & medium & - & 3 \\
3.3 & - & - & - & NSDSECPLTHDGYCLHDGVCQFIKDMQEYACSCVVGYIGERCKHRDLKFWGLR & high & - & 3 \\
2 & - & - & - & NSYPECPPSYEKWYCLHGTCRFVQELKDAACICHKGYMGSRCEFLTLDW & high & - & 3 \\
3.2 & - & - & - & NTIQCPSSHDTYCLYGGACFYFKVLQDYACQCAEGYVGERCDNSDLSWDEFR & high & - & 3 \\
3.2 & - & - & - & NTTSDGTGCPLSHDSYCLNGGVCLYYTPMKQWACDCVSGYTGERCQFKDL & medium & - & 3 \\
3.1 & - & - & - & PARDVPEPCLSGNFPLPELYSAPCPKELEHYCIHGTCRYLEGLNKPSCRCQKGYVGSRCEYYEL & none & - & 3 \\
3.2 & - & - & - & PARTTGHLARPCPMPESYCLNGGTCSFYSTIGKYVCQCAEGFKGQRCENQD & low & - & 3 \\
3.2 & - & - & - & PCPMEDFCLNGAATFFCLKGYCLRVKELQSFACQCLKGYDGERCGIQKMK & medium & - & 3 \\
3.1 & - & - & - & PEDAERGEGCPPSYCLNGGQCLFFETVQNYACNCVAGYIGERCEHSDLEWWEL & medium & - & 3 \\
2 & - & - & - & PETAAELREVSSGAARCPPTYDSFCLHQGICVFYPEVDKYACNCVKGYMGQRCQFSDLEWW & medium & - & 3 \\
3.1 & - & - & - & PPTEVEDGQGRIPECPESYCLNEGVCFHIPEIRAYACHCLKGYMGERCEFSDL & medium & - & 3 \\
3.2 & - & - & - & PSDSSLSEPCPVDGYCLNGGLCLYFEAIGKFVCKCAEGFLGERCERRHLLWYEL & high & - & 3 \\
3.2 & - & - & - & PTKNGHAEPCPMPDSYCLNGGICIYFSIIQEYVCNCVTGFLGPRCNHLQFID & high & - & 3 \\
2 & - & - & - & SGPLRSPEPATDFPGRGLPCPVDHYCLNGGRCKFLEELQEYACECVPGYIGERCQYDDLTWWA & low & - & 3 \\
3.3 & - & - & - & SGSTECPPSQDSYCLHNGVCMYIEPIRAYACKCVVGFTGDRCQFRDLRWRE & high & - & 3 \\
3.2 & - & - & - & SNASPCPMPDPCTNGETVFCLNGGTCLFYELLQEHACKCKPGFTGARCENKEI & low & - & 3 \\
3.2 & - & - & - & SNGTPCPMPDAHDSYCLNGGVCFHYEQLKPVACICHKGYTGERCQHNSTP & high & - & 3 \\
3.3 & - & - & - & SNSIEDCPSAHDSYCLYGGVCFYIPNMSKFACKCLVGYMGERCQFFDWWYSQ & high & - & 3 \\
3.3 & - & - & - & SNSVESCPSSHETFCLYDGVCFHIKDLKNYACNCLTGYMGKRCQYSDLEWWE & high & - & 3 \\
2 & - & - & - & SNTVERCPSAHDTYCLYGGVCFYIPELQQYACNCVPGYVGERCQYSDLEWWE & none & - & 3 \\
3.1 & - & - & - & SPCPEEYEHLNGYCIHGTCRFVQDLEKVACRCLKGYTGERCEHFTMLPWDS & high & - & 3 \\
2 & - & - & - & SPETDTINGLAKCPEEYIHFCIHGVCKFIQEQETPSCRCNKGYNGSRCE & high & - & 3 \\
3.2 & - & - & - & SPGPATLRSSSTTHSSCPVTDFCLNGGTCSYYETIGEMVCQCADGYKGRRCSNRD & medium & - & 3 \\
3.1 & - & - & - & SPGVETIAECPEKYAKCVEGYCRHMELLRQYACDCLAGYTGERCESNDLTWWADW & none & - & 3 \\
3.3 & - & - & - & STVDSCPSAHDSYCLYGGVCFHIAELKGYACNCVIGYVGERCQHRDLRWWDLR & high & - & 3 \\
3.2 & - & - & - & TNFDCPESHYEGYCLHGGVCMHIKAVKDHACECIRGYSGEKCQKQDLYTLNW & medium & - & 3 \\
2 & - & - & - & TNTVDTCPSTYCLHEGICFYFSIMQSFACNCVRGYLGDRCQFSDLEWWE & high & - & 3 \\
3.1 & - & - & - & TPCNGEFSLNSYCLYGDLCYFMPEVEAYACERKWKCLCQSGYLGERCSYSDLEWWE & high & - & 3 \\
3.2 & - & - & - & TPTESSTSSVCPLDAGFCLNGGICLYYPEVQRFACKCAEGYFGERCQYEDLEWWE & high & - & 3 \\
3.2 & - & - & - & TSNAGPVCPLDSYCLNGGICIYFSEVQAYACKCVVGYVGDRCQHSDMEWWE & high & - & 3 \\
3.1 & - & - & - & TSVPSDARGLARRCPPSYCLNEGACLFIEAVNRYACKCLPNFIGERCQFSDLEWWE & medium & - & 3 \\
3.1 & - & - & - & TVPPELSRSGNCVKGFCLHGTCKYIEQLRAVACNCLKGYEGPRCQHKDLRWWEL & medium & - & 3 \\
\end{longtable}
\end{landscape}

\begin{figure}
    \centering
    \includegraphics[width=1\linewidth]{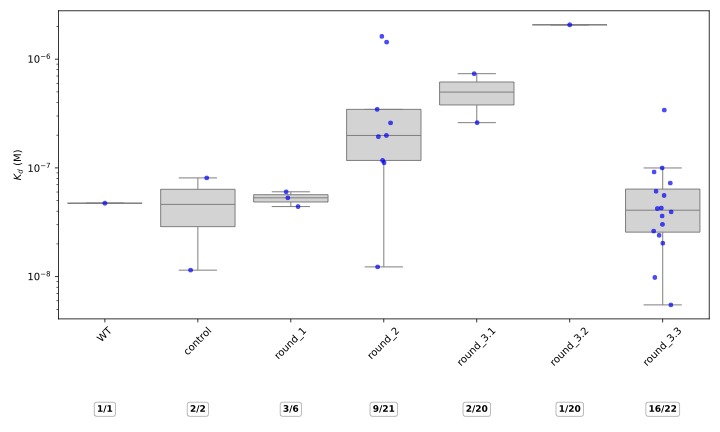}
    \caption{All the experiments carried and compared with reported also the number of tested elements with the number of successful binders for each round. Round 2 results from the alignment of the Round 1 model (FT model) using ProtRL (GitHub release 1.0.0). Rounds 3.1 and 3.2 are based on the alignment of the model that generated sequences in Round 2. Round 3.3 uses ProtRL (GitHub release 2.0.0), starting with the Round 1 (SFT) model, aligning the model once with merged experimental data obtained from Rounds 2, 3.1, and 3.2 and tested controls. The final experiment resulted not only in better binders in average, in achieving best binder reported in literature, but also remarkable ratio of successful binders over tested ($72\%$)}
    \label{fig:comparison_rounds}
\end{figure}

\begin{figure}[h]
  \centering
  \includegraphics[width=\textwidth,page=1]{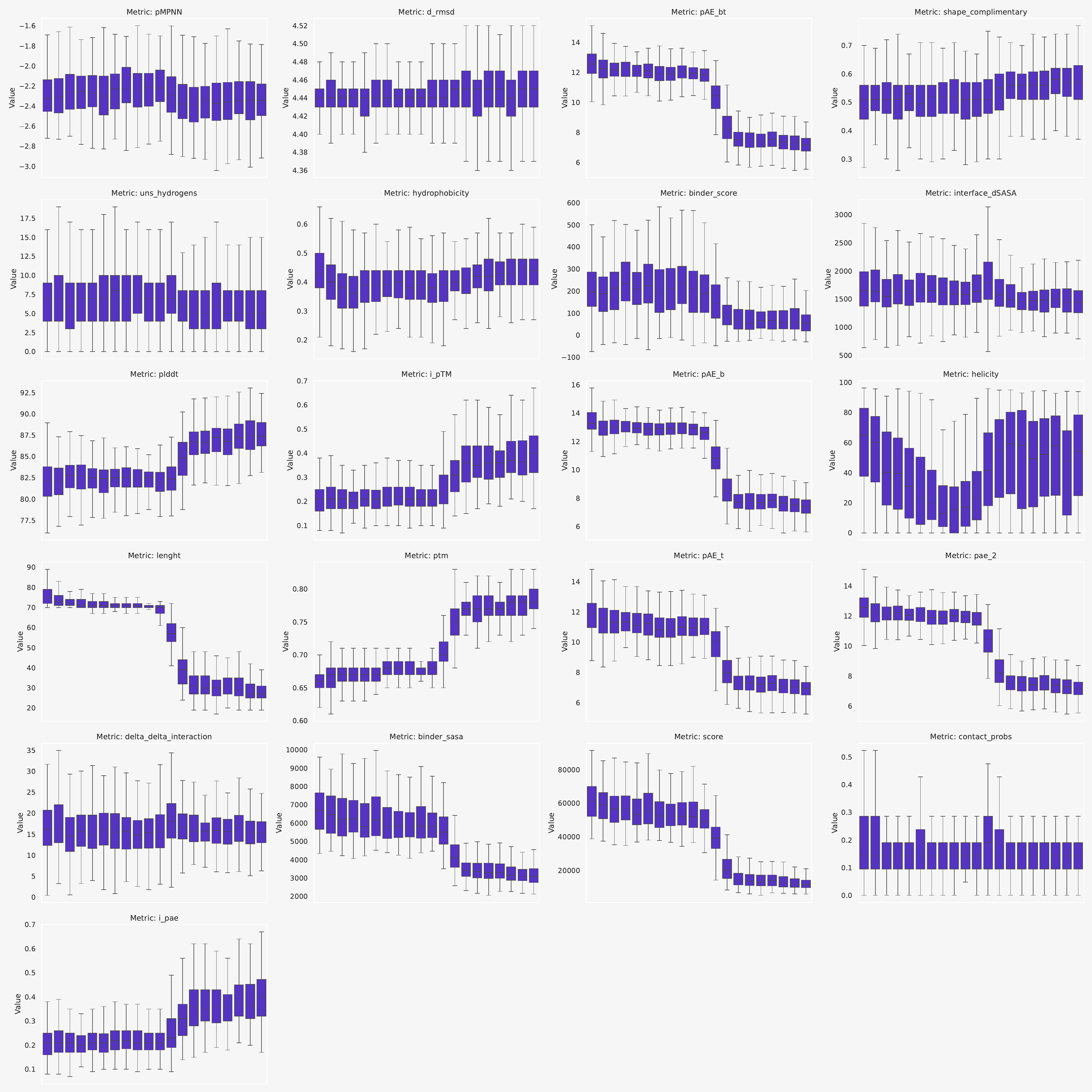}
  \caption{\textbf{Multi-objective optimization of CD80 using ProtRL.} Optimization was initiated from a fine-tuned autoregressive protein language model. Shown is the progression of the optimization process across successive iterations, with the corresponding changes in the monitored metrics.}
  \label{fig:SI_cd80_protwrap}
\end{figure}

\begin{figure}
    \centering
    \includegraphics[width=0.7\linewidth]{figures/ProtWrap_nipah.png}
    \caption{\textbf{Multi-objective optimization of binding sequences for Nipah virus glycoprotein G using ProtRL.} Optimization was initiated from a fine-tuned autoregressive protein language model using the REINFORCE algorithm implemented in ProtRL (v2.0.0). Shown is the progression of the monitored metrics across successive iterations. Dashed lines indicate the thresholds reported in Table~S.\ref{tab:feature_auc}. After iteration 8, performance reverses and the model undergoes collapse consistent with catastrophic forgetting. This behaviour underscores the importance of closely monitoring the alignment process and highlights the sensitivity of successful optimization to hyperparameter selection.}
    \label{fig:SI_nipah_protwrap}
\end{figure}

\begin{figure}[h]
  \centering
  \includegraphics[width=\textwidth,page=1]{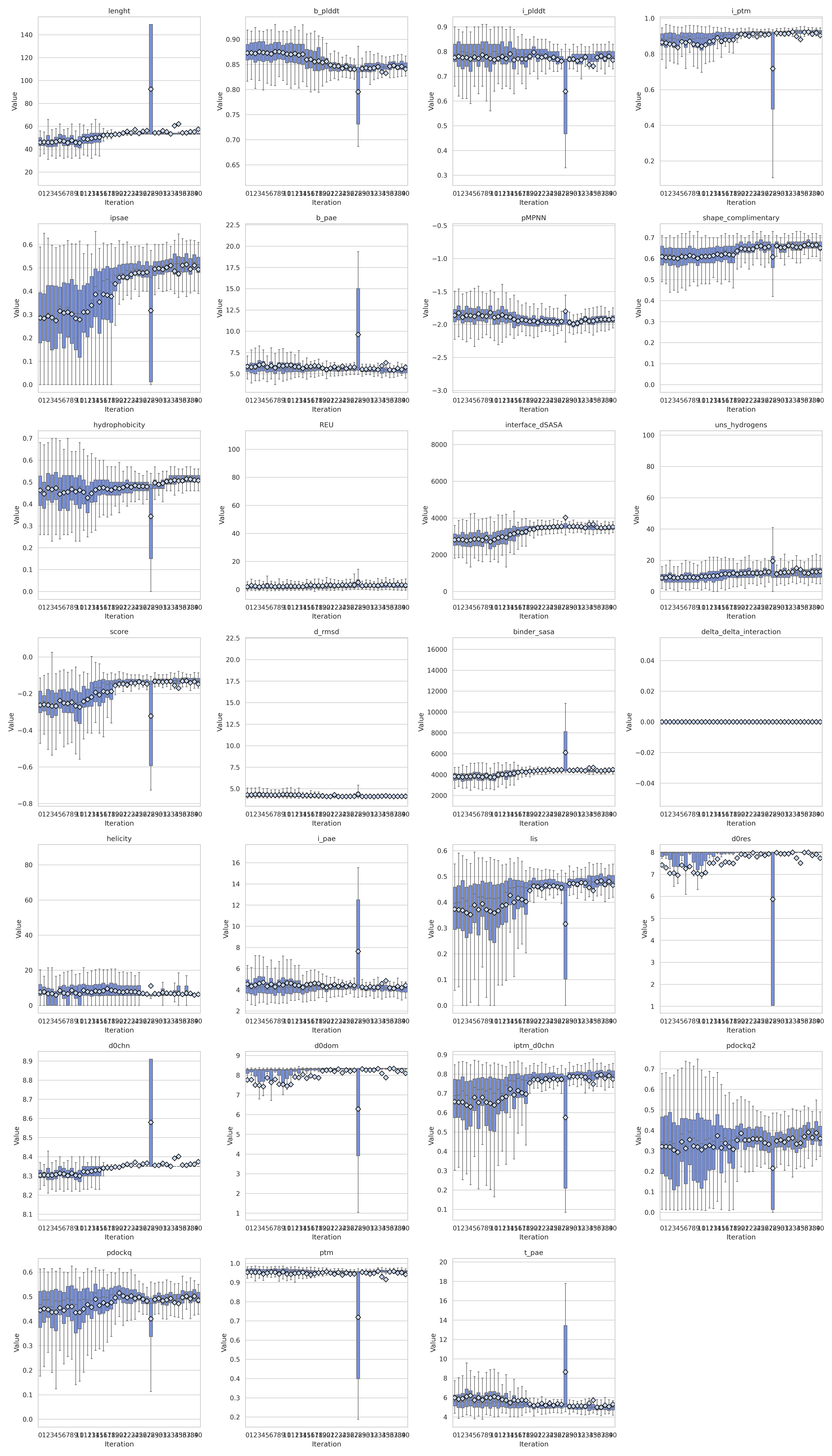}
  \caption{\textbf{Multi-objective optimization of EGF using ProtRL.} Optimization was initiated from a fine-tuned autoregressive protein language model. Shown is the progression of the optimization process across successive iterations, with the corresponding changes in the monitored metrics.}
  \label{fig:SI_egf_protwrap}
\end{figure}

\begin{table}[h]
\centering
\caption{Hyperparameters used for ProtRL training unless otherwise specified.}
\label{Table:GRPO_hyperparameters}
\begin{tabular}{lc}
\toprule
Hyperparameter & Value \\
\midrule
$\beta$ & 0.01 or 0.1 \\
Seed & 42 \\
Learning rate & $1\times10^{-6}$ or $2\times10^{-7}$\\
Batch size & 15 \\
Epochs & 1 \\
Train/Test split & 0.2 \\
Adam betas & (0.9, 0.98) \\
$\epsilon$ & $1\times10^{-8}$ \\
Weight decay & 0.1 \\
\bottomrule
\end{tabular}
\end{table}

\end{document}